\documentclass[prc,reprint,onecolumn,amsmath,amssymb,superscriptaddress]{revtex4-1}
\usepackage{graphicx}

\newcommand{\be}{\begin{equation}}
\newcommand{\ee}{\end{equation}}
\newcommand{\beq}{\begin{eqnarray}}
\newcommand{\eeq}{\end{eqnarray}}
\newcommand{\ba}{\begin{array}}
\newcommand{\ea}{\end{array}}

\usepackage{color}

\begin{document}

\title{Cross Section for $\gamma n\rightarrow\pi^0n$ measured at
	Mainz/A2}

\date{\today}

\author{\mbox{W.~J.~Briscoe}}
\affiliation{Institute for Nuclear Studies, Department of Physics, The  George 
	Washington University, Washington DC 20052, USA}

\author{\mbox{M.~Had\v{z}imehmedovi\'{c}}}
\affiliation{University of Tuzla, Faculty of Natural Sciences and Mathematics, \\ 
	Univerzitetska 4, 75000 Tuzla, Bosnia and Herzegovina}

\author{\mbox{A.~E.~Kudryavtsev}}
\affiliation{National Research Centre ``Kurchatov Institute", Institute for 
	Theoretical and Experimental Physics (ITEP), Moscow 117218, Russia}
\affiliation{Institute for Nuclear Studies, Department of Physics, The George 
	Washington University, Washington DC 20052, USA}

\author{\mbox{V.~V.~Kulikov}}
\affiliation{National Research Centre ``Kurchatov Institute", Institute for 
	Theoretical and Experimental Physics (ITEP), Moscow 117218, Russia}

\author{\mbox{M.~A.~Martemianov}}
\affiliation{National Research Centre ``Kurchatov Institute", Institute for 
	Theoretical and Experimental Physics (ITEP), Moscow 117218, Russia}

\author{\mbox{I.~I.~Strakovsky}}
\altaffiliation{Corresponding author; \texttt{igor@gwu.edu}}
\affiliation{Institute for Nuclear Studies, Department of Physics, The George 
	Washington University, Washington DC 20052, USA}

\author{\mbox{A.~\v{S}varc}}
\affiliation{Rudjer Bo\v{s}kovi\'{c} Institute, Bijeni\v{c}ka cesta 54, 10002 
	Zagreb, Croatia}
\affiliation{Tesla Biotech, Mandlova 7, 10002 Zagreb, Croatia}

\author{\mbox{V.~E.~Tarasov}}
\affiliation{National Research Centre ``Kurchatov Institute", Institute for
        Theoretical and Experimental Physics (ITEP), Moscow 117218, Russia}

\author{\mbox{R.~L.~Workman}}
\affiliation{Institute for Nuclear Studies, Department of Physics, The George 
	Washington University, Washington DC 20052, USA}

\author{\mbox{S.~Abt}}
\affiliation{Department of Physics, University of Basel, Ch-4056 Basel, 
	Switzerland}

\author{\mbox{P.~Achenbach}}
\affiliation{Institut f\"ur Kernphysik, University of Mainz, D-55099 Mainz, Germany}

\author{\mbox{C.~S.~Akondi}}
\affiliation{Kent State University, Kent, OH 44242, USA}

\author{\mbox{F.~Afzal}}
\affiliation{Helmholtz-Institut f\"ur Strahlen- und Kernphysik, University Bonn, 
	D-53115 Bonn, Germany}

\author{\mbox{P.~Aguar-Bartolom\'e}}
\affiliation{Institut f\"ur Kernphysik, University of Mainz, D-55099 Mainz, Germany}

\author{\mbox{Z.~Ahmed}}
\affiliation{University of Regina, Regina, SK S4S-0A2 Canada}

\author{\mbox{J.~R.~M.~Annand}}
\affiliation{SUPA School of Physics and Astronomy, University of Glasgow, Glasgow, 
	G12 8QQ, UK}

\author{\mbox{H.~J.~Arends}}
\affiliation{Institut f\"ur Kernphysik, University of Mainz, D-55099 Mainz, Germany}

\author{\mbox{K.~Bantawa}}
\affiliation{Kent State University, Kent, OH 44242, USA}

\author{\mbox{M.~Bashkanov}}
\affiliation{Department of Physics, University of York, Heslington, York, Y010 5DD, UK}

\author{\mbox{R.~Beck}}
\affiliation{Helmholtz-Institut f\"ur Strahlen- und Kernphysik, University Bonn, D-53115 
	Bonn, Germany}

\author{\mbox{M.~Biroth}}
\affiliation{Institut f\"ur Kernphysik, University of Mainz, D-55099 Mainz, Germany}

\author{\mbox{N.~Borisov}}
\affiliation{Joint Institute for Nuclear Research, RU-141980 Dubna, Russia}

\author{\mbox{A.~Braghieri}}
\affiliation{INFN Sezione di Pavia, I-27100 Pavia, Pavia, Italy}

\author{\mbox{S.~A.~Bulychjov}}
\affiliation{National Research Centre ``Kurchatov Institute", Institute for Theoretical 
	and Experimental Physics (ITEP), Moscow 117218, Russia}

\author{\mbox{F.~Cividini}}
\affiliation{Institut f\"ur Kernphysik, University of Mainz, D-55099 Mainz, Germany}

\author{\mbox{C.~Collicott}}
\affiliation{Department of Astronomy and Physics, Saint Mary's University, E4L1E6 
	Halifax, Canada}

\author{\mbox{S.~Costanza}}
\affiliation{Dipartimento di Fisica, Universit\`a di Pavia, I-27100 Pavia, Italy}
\affiliation{INFN Sezione di Pavia, I-27100 Pavia, Pavia, Italy}

\author{\mbox{A.~Denig}}
\affiliation{Institut f\"ur Kernphysik, University of Mainz, D-55099 Mainz, Germany}

\author{\mbox{E.~J.~Downie}}
\affiliation{Institute for Nuclear Studies, Department of Physics, The
        George Washington University, Washington DC 20052, USA}

\author{\mbox{P.~Drexler}}
\affiliation{Institut f\"ur Kernphysik, University of Mainz, D-55099 Mainz, Germany}

\author{\mbox{S.~Fegan}}
\affiliation{Institute for Nuclear Studies, Department of Physics, The
        George Washington University, Washington DC 20052, USA}

\author{\mbox{M.~I.~Ferretti Bondy}}
\affiliation{Institut f\"ur Kernphysik, University of Mainz, D-55099 Mainz, Germany}

\author{\mbox{S.~Gardner}}
\affiliation{SUPA School of Physics and Astronomy, University of Glasgow, Glasgow,
        G12 8QQ, UK}

\author{\mbox{D.~Ghosal}}
\affiliation{Department of Physics, University of Basel, Ch-4056 Basel,
        Switzerland}

\author{\mbox{D.~I.~Glazier}}
\affiliation{SUPA School of Physics and Astronomy, University of Glasgow, Glasgow, 
	G12 8QQ, UK}

\author{\mbox{I.~Gorodnov}}
\affiliation{Joint Institute for Nuclear Research, RU-141980 Dubna, Russia}

\author{\mbox{W.~Gradl}}
\affiliation{Institut f\"ur Kernphysik, University of Mainz, D-55099 Mainz, Germany}

\author{\mbox{M.~G\"unther}}
\affiliation{Department of Physics, University of Basel, Ch-4056 Basel,
        Switzerland}

\author{\mbox{D.~Gurevich}}
\affiliation{Institute for Nuclear Research, RU-125047 Moscow, Russia}

\author{\mbox{L.~Heijkenskj{\"o}ld}}
\affiliation{Institut f\"ur Kernphysik, University of Mainz, D-55099 Mainz, Germany}

\author{\mbox{D.~Hornidge}}
\affiliation{Mount Allison University, Sackville, New Brunswick E4L1E6, Canada}

\author{\mbox{G.~M.~Huber}}
\affiliation{University of Regina, Regina, SK S4S-0A2 Canada}

\author{\mbox{M.~K\"aser}}
\affiliation{Department of Physics, University of Basel, Ch-4056 Basel,
        Switzerland}

\author{\mbox{V.~L.~Kashevarov}}
\affiliation{Institut f\"ur Kernphysik, University of Mainz, D-55099 Mainz, Germany}
\affiliation{Joint Institute for Nuclear Research, RU-141980 Dubna, Russia}

\author{\mbox{S.~Kay}}
\affiliation{University of Regina, Regina, SK S4S-0A2 Canada}

\author{\mbox{M.~Korolija}}
\affiliation{Rudjer Bo\v{s}kovi\'{c} Institute, Bijeni\v{c}ka cesta 54, 10002
        Zagreb, Croatia}
\affiliation{Tesla Biotech, Mandlova 7, 10002 Zagreb, Croatia}

\author{\mbox{B.~Krusche}}
\affiliation{Department of Physics, University of Basel, Ch-4056 Basel,
        Switzerland}

\author{\mbox{A.~Lazarev}}
\affiliation{Joint Institute for Nuclear Research, RU-141980 Dubna, Russia}

\author{\mbox{K.~Livingston}}
\affiliation{SUPA School of Physics and Astronomy, University of Glasgow, Glasgow, 
	G12 8QQ, UK}

\author{\mbox{S.~Lutterer}}
\affiliation{Department of Physics, University of Basel, Ch-4056 Basel,
        Switzerland}

\author{\mbox{I.~J.~D.~MacGregor}}
\affiliation{SUPA School of Physics and Astronomy, University of Glasgow, Glasgow, 
	G12 8QQ, UK}

\author{\mbox{R.~Macrae}}
\affiliation{SUPA School of Physics and Astronomy, University of Glasgow, Glasgow, 
	G12 8QQ, UK}

\author{\mbox{D.~M.~Manley}}
\affiliation{Kent State University, Kent, OH 44242, USA}

\author{\mbox{P.~P.~Martel}}
\affiliation{Institut f\"ur Kernphysik, University of Mainz, D-55099 Mainz, Germany}

\author{\mbox{J.~C.~McGeorge}}
\affiliation{SUPA School of Physics and Astronomy, University of Glasgow, Glasgow, 
	G12 8QQ, UK}

\author{\mbox{D.~G.~Middleton}}
\affiliation{Institut f\"ur Kernphysik, University of Mainz, D-55099 Mainz, Germany}
\affiliation{Mount Allison University, Sackville, New Brunswick E4L1E6, Canada}

\author{\mbox{R.~Miskimen}}
\affiliation{University of Massachusetts, Amherst, MA 01003, USA}

\author{\mbox{E.~Mornacchi}}
\affiliation{Institut f\"ur Kernphysik, University of Mainz, D-55099 Mainz, Germany}

\author{\mbox{A.~Mushkarenkov}}
\affiliation{INFN Sezione di Pavia, I-27100 Pavia, Pavia, Italy}
\affiliation{University of Massachusetts, Amherst, MA 01003, USA}

\author{\mbox{C.~Mullen}}
\affiliation{SUPA School of Physics and Astronomy, University of Glasgow, Glasgow, 
	G12 8QQ, UK}

\author{\mbox{A.~Neganov}}
\affiliation{Joint Institute for Nuclear Research, RU-141980 Dubna, Russia}

\author{\mbox{A.~Neiser}}
\affiliation{Institut f\"ur Kernphysik, University of Mainz, D-55099 Mainz, Germany}

\author{\mbox{M.~Ostrick}}
\affiliation{Institut f\"ur Kernphysik, University of Mainz, D-55099 Mainz, Germany}

\author{\mbox{P.~B.~Otte}}
\affiliation{Institut f\"ur Kernphysik, University of Mainz, D-55099 Mainz, Germany}

\author{\mbox{H.~Osmanovi\'{c}}}
\affiliation{University of Tuzla, Faculty of Natural Sciences and Mathematics, \\ 
	Univerzitetska 4, 75000 Tuzla, Bosnia and Herzegovina}

\author{\mbox{R.~Omerovi\'{c}}}
\affiliation{University of Tuzla, Faculty of Natural Sciences and Mathematics, \\ 
	Univerzitetska 4, 75000 Tuzla, Bosnia and Herzegovina}

\author{\mbox{B.~Oussena}}
\affiliation{Institut f\"ur Kernphysik, University of Mainz, D-55099 Mainz, Germany}
\affiliation{Institute for Nuclear Studies, Department of Physics, The
        George Washington University, Washington DC 20052, USA}

\author{\mbox{D.~Paudyal}}
\affiliation{University of Regina, Regina, SK S4S-0A2 Canada}

\author{\mbox{P.~Pedroni}}
\affiliation{INFN Sezione di Pavia, I-27100 Pavia, Pavia, Italy}

\author{\mbox{A.~Powell}}
\affiliation{SUPA School of Physics and Astronomy, University of Glasgow, Glasgow,
        G12 8QQ, UK}

\author{\mbox{S.~N.~Prakhov}}
\affiliation{Institut f\"ur Kernphysik, University of Mainz, D-55099 Mainz, Germany}
\affiliation{University of California Los Angeles, Los Angeles, CA 90095, USA}

\author{\mbox{G.~Ron}}
\affiliation{Racah Institute of Physics, Hebrew University of Jerusalem, Jerusalem 
	91904, Israel}
\affiliation{Institute for Nuclear Studies, Department of Physics, The  George
        Washington University, Washington DC 20052, USA}

\author{\mbox{T.~Rostomyan}}
\affiliation{Department of Physics, University of Basel, Ch-4056 Basel,
        Switzerland}

\author{\mbox{A.~Sarty}}
\affiliation{Department of Astronomy and Physics, Saint Mary's University, E4L1E6 
	Halifax, Canada}

\author{\mbox{C.~Sfienti}}
\affiliation{Institut f\"ur Kernphysik, University of Mainz, D-55099 Mainz, Germany}

\author{\mbox{V.~Sokhoyan}}
\affiliation{Institut f\"ur Kernphysik, University of Mainz, D-55099 Mainz, Germany}

\author{\mbox{K.~Spieker}}
\affiliation{Helmholtz-Institut f\"ur Strahlen- und Kernphysik, University Bonn, D-53115 
	Bonn, Germany}

\author{\mbox{J.~Stahov}}
\affiliation{University of Tuzla, Faculty of Natural Sciences and Mathematics, \\ 
	Univerzitetska 4, 75000 Tuzla, Bosnia and Herzegovina}

\author{\mbox{O.~Steffen}}
\affiliation{Institut f\"ur Kernphysik, University of Mainz, D-55099 Mainz, Germany}

\author{\mbox{I.~Supek}}
\affiliation{Rudjer Bo\v{s}kovi\'{c} Institute, Bijeni\v{c}ka cesta 54, 10002
        Zagreb, Croatia}
\affiliation{Tesla Biotech, Mandlova 7, 10002 Zagreb, Croatia}

\author{\mbox{A.~Thiel}}
\affiliation{Helmholtz-Institut f\"ur Strahlen- und Kernphysik, University Bonn, D-53115
        Bonn, Germany}
\affiliation{SUPA School of Physics and Astronomy, University of Glasgow, Glasgow,
        G12 8QQ, UK}

\author{\mbox{M.~Thiel}}
\affiliation{Institut f\"ur Kernphysik, University of Mainz, D-55099 Mainz, Germany}

\author{\mbox{A.~Thomas}}
\affiliation{Institut f\"ur Kernphysik, University of Mainz, D-55099 Mainz, Germany}

\author{\mbox{L.~Tiator}}
\affiliation{Institut f\"ur Kernphysik, University of Mainz, D-55099 Mainz, Germany}

\author{\mbox{M.~Unverzagt}}
\affiliation{Institut f\"ur Kernphysik, University of Mainz, D-55099 Mainz, Germany}

\author{\mbox{Yu.~A.~Usov}}
\affiliation{Joint Institute for Nuclear Research, RU-141980 Dubna, Russia}

\author{\mbox{N.~K.~Walford}}
\affiliation{Department of Physics, University of Basel, Ch-4056 Basel, Switzerland}

\author{\mbox{D.~P.~Watts}}
\affiliation{Department of Physics, University of York, Heslington, York, Y010 5DD, UK}

\author{\mbox{S.~Wagner}}
\affiliation{Institut f\"ur Kernphysik, University of Mainz, D-55099 Mainz, Germany}

\author{\mbox{D.~Werthm\"uller}}
\affiliation{Department of Physics, University of York, Heslington, York, Y010 5DD, UK}

\author{\mbox{J.~Wettig}}
\affiliation{Institut f\"ur Kernphysik, University of Mainz, D-55099 Mainz, Germany}

\author{\mbox{M.~Wolfes}}
\affiliation{Institut f\"ur Kernphysik, University of Mainz, D-55099 Mainz, Germany}

\author{\mbox{N.~Zachariou}}
\affiliation{Department of Physics, University of York, Heslington, York, Y010 5DD, UK}

\collaboration{A2 Collaboration at MAMI}
\noaffiliation

\begin{abstract}
The $\gamma n\rightarrow\pi^0n$ differential cross section
evaluated for 27 energy bins span the photon-energy range 
290--813~MeV ($W$ = 1.195 -- 1.553~GeV) and the pion c.m. 
polar production angles, ranging from 18$^\circ$ to 
162$^\circ$, making use of model-dependent nuclear
corrections to extract $\pi^0$ production data on the 
neutron from measurements on the deuteron target. 
Additionally, the total photoabsorption cross section was 
measured. The tagged photon beam produced by the 883-MeV 
electron beam of the Mainz Microtron MAMI was used for 
the $\pi^0$-meson production. Our accumulation of $3.6\times 
10^6$ $\gamma n\rightarrow\pi^0n$ events allowed a detailed 
study of the reaction dynamics. Our data are in reasonable 
agreement with previous A2 measurements and extend them to 
lower energies.  The data are compared to predictions of 
previous SAID, MAID, and BnGa partial-wave analyses and to 
the latest SAID fit MA19 that included our data. Selected 
photon decay amplitudes $N^\ast\rightarrow\gamma n$ at the 
resonance poles are determined for the first time.
\end{abstract}

\maketitle

\section{Introduction}
\label{sec:intro}

The $N^\ast$ and $\rm\Delta^\ast$ families of nucleon resonances 
have many well-established members~\cite{Tanabashi:2018oca}, 
several of which overlap, having very similar masses and widths 
but different J$^{\rm P}$ spin-parity values. There are two closely 
spaced states above the famous $\rm\Delta$(1232)$3/2^+$ resonance: 
N(1520)$3/2^-$ and N(1535)$1/2^-$.
\begin{table}[htb!]

\centering \protect\caption{Breit-Wigner mass and full width (in MeV) 
	with proton $A_{3/2}(p)$ and $A_{1/2}(p)$, and with neutron 
	$A_{3/2}(n)$ and $A_{1/2}(n)$ BW photon decay 
	amplitudes (in $\rm (GeV)^{-1/2}\times 10^{-3}$) from the 
	PDG2018~\protect\cite{Tanabashi:2018oca} covering the energy
	range of the A2 experiment.}
\vspace{2mm}
{\begin{tabular}{|c|c|c|c|c|c|c|} \hline
Resonance           & $M$         & $\Gamma$   & $A_{3/2}(p)$ & $A_{1/2}(p)$ & $A_{3/2}(n)$ & $A_{1/2}(n)$ \tabularnewline
\hline
$\rm\Delta(1232)3/2^+$&1232$\pm$2 & 117$\pm$3  & -255$\pm$7   &-135$\pm$6    &              &              \tabularnewline
N(1440)$1/2^+$      & 1440$\pm$30 & 350$\pm$100&              & -65$\pm$15   &              & +45$\pm$10   \tabularnewline
N(1520)$3/2^-$      & 1515$\pm$15 & 110$\pm$10 & +140$\pm$5   & -25$\pm$8    & -115$\pm$8   & -50$\pm$8    \tabularnewline
N(1535)$1/2^-$      & 1530$\pm$15 & 150$\pm$25 &              & +105$\pm$15  &              & -75$\pm$20   \tabularnewline
\hline
\end{tabular}} \label{tab:tbl1}
\end{table}

One critical issue in the study of meson photoproduction on the
nucleon comes from isospin. While isospin can change at the photon
vertex, it must be conserved at the final hadronic vertex. The 
isospin amplitudes for the $\gamma N \to \pi N$ reactions are 
decomposed into three distinct $I=1/2$ (proton and neutron) and
$I=3/2$ isospin components, $A_{\gamma,\pi^0 p/n} = \pm 
A_{p/n}^{I=1/2} +  \frac{2}{3}A^{I=3/2}$ and $A_{\gamma,\pi^\pm} = 
\sqrt{2}(A_{p/n}^{I=1/2} \mp \frac{1}{3}A^{I=3/2})$ (see 
Ref.~\cite{Drechsel:1992pn}).  This expression indicates that the 
$I = 3/2$ multipoles can be entirely determined from proton target 
data.  However, measurements from datasets with both neutron and 
proton targets are required to determine the isospin $I = 1/2$ 
amplitudes and to separate the $\gamma pN^\ast$ and $\gamma 
nN^\ast$ photon couplings. Only with good data on both proton and 
neutron targets one can hope to disentangle the isoscalar and 
isovector electromagnetic (EM) couplings of the various $N^\ast$ 
and $\rm\Delta^\ast$ resonances~\cite{Watson:1954uc,Walker:1968xu}, 
as well as the isospin properties of the non-resonant background 
amplitudes. The lack of $\gamma n\rightarrow\pi^-p$ and $\gamma 
n\rightarrow\pi^0n$ data~\cite{SAID} does not allow us to be as 
confident about the determination of neutron couplings compared 
to those of the proton.  Some of the $N^\ast$-baryons 
($N(1520)3/2^-$, for instance) have stronger EM couplings to the 
neutron relative to the proton, while others (for instance the 
nearby $N(1535)1/2^-$) have weaker EM couplings to the 
neutron relative to the proton. However, the resonance parameters 
of both these states are very uncertain, see Table~\ref{tab:tbl1}.

In the SAID $\pi N$ partial-wave analysis (PWA), one can determine 
$\pi N$ amplitudes by fitting the $\pi N$ elastic data (up to $W$ = 
2.5~GeV)~\cite{Arndt:2006bf,Workman:2012hx}. Resonances are then found 
through a search for poles in the complex energy plane. The SAID group 
considers mainly poles which are not far away from the physical axis. 
It is important to emphasize that these resonances are not put in by 
hand, contrary to the Breit-Wigner (BW) parametrization. The poles 
arise, in a sense, dynamically as a result of the enforced (quasi-) 
two-body unitarity cuts and the fit to the observable on the real 
energy axis. The photoproduction multipoles can be parametrized using 
a form containing the Born terms (with no free parameters), 
phenomenological pieces maintaining the correct threshold behavior, 
and Migdal-Watson's theorem~\cite{Migdal:1955,Watson:1952ji} below 
the two-pion production threshold. The $\pi N$ matrix connects each 
multipole to structure found in the elastic scattering analysis. The 
parametrization above the two-pion production threshold is based on 
a unitary $K$-matrix approach, with no strong constraints on the 
energy dependence apart from correct threshold properties.

Knowledge of the $N^\ast$ and $\rm\Delta^\ast$ resonance decay 
amplitudes into nucleons and photons is largely restricted to charged 
states.  Apart from the lower-energy inverse reaction $\pi^-p\rightarrow
\gamma n$ measurements, the extraction of the two-body $\gamma 
n\rightarrow\pi^-p$ and $\gamma n\rightarrow\pi^0n$ observables 
requires the use of a model-dependent nuclear correction, which 
mainly comes from final-state interaction (FSI) effects within the 
deuteron. Most $\gamma n$ data are unpolarized and cover fairly 
narrow energy ranges.  Of these, only about 500 $\pi^0n$ 
measurement data points exist, spanning the limited nucleon 
resonance region~\cite{SAID}.

A FSI correction factor was defined as the ratio between a sum of 
leading diagrams and an impulse-approximation (IA) that the GWU-ITEP 
group then applied to the experimental $\gamma d$ data to get a two-body 
cross section for $\gamma n\rightarrow\pi^-p$~\cite{Tarasov:2011ec} 
and $\gamma n\rightarrow\pi^0n$~\cite{Tarasov:2015sta}. The GWU SAID 
phenomenological amplitudes for $\pi N$ and $NN$ elastic scattering  
and $\gamma N\rightarrow\pi N$ were used as inputs to calculate the 
leading diagrams for the GWU-ITEP FSI code.  The full Bonn potential 
was used for the deuteron description. Recently, the GWU-ITEP group 
determined $\gamma n\rightarrow\pi^-p$ differential cross sections 
from $\gamma d\rightarrow\pi^-pp$ measurements made by the 
CLAS~\cite{Chen:2012yv,Mattione:2017fxc} and 
MAMI/A2~\cite{Briscoe:2012ni} Collaborations.  In this way, we 
succeeded in the first determination of neutron couplings at the 
pole positions for a number of baryons, such as $N(1440)1/2^+$, 
$N(1535)1/2^-$, $N(1650)1/2^-$, and $N(1720)3/2^+$, 
significantly improving the world data~\cite{Mattione:2017fxc}.

The $\gamma n\rightarrow\pi^0n$ measurement on the deuteron target 
is much more complicated than $\gamma n\rightarrow\pi^-p$ because 
the $\pi^0$ can come from both $\gamma n$ and $\gamma p$ initial 
states. The GWU-ITEP studies have shown that photoproduction cross 
sections off the protons and neutrons are generally not 
equal~\cite{Tarasov:2015sta}: 
\begin{eqnarray}
        A(\gamma p\rightarrow\pi^0p) = A_v + A_s, ~~~~~
        A(\gamma n\rightarrow\pi^0n) = A_v - A_s,
        \label{eq:iso}
\end{eqnarray}\noindent
where $A_v$ and $A_s$ are isovector and isoscalar amplitudes,
respectively, and $A_s\neq 0$.  However, in the special case, in the 
region of the $\rm\Delta(1232)3/2^+$ and $A_s = 0$, the FSI corrections 
for $\gamma p\rightarrow\pi^0p$ and $\gamma n\rightarrow\pi^0n$ cross 
sections are equal due to the isospin structure of the $\gamma 
N\rightarrow\pi N$ amplitudes.

Recently, the A2 Collaboration at MAMI published high-quality 
unpolarized and polarized measurements for $\pi^0$ photoproduction 
on a proton target below $W$ = 2~GeV~\cite{Gardner:2016irh,Annand:2016ppc,
Schumann:2015ypa,Adlarson:2015byy,Sikora:2013vfa,Hornidge:2012ca} while 
one study was carried out for $\pi^0$ photoproduction on the 
neutron~\cite{Dieterle:2018adj,Dieterle:2017myg}.  This last study focuses 
on neutral pion photoproduction off the neutron using a deuteron target.

In the present paper new, precise, $\gamma n\rightarrow\pi^0n$ 
differential cross sections for $E$ = 290 to 813~MeV in laboratory 
photon energy, corresponding to center-of-mass (c.m.) energy range from 
$W$ = 1.195 to 1.553~GeV, are reported.  Pion c.m.\ polar production 
angles, ranging from $\theta$ = 18$^\circ$ to 162$^\circ$, have been 
measured by the A2 Collaboration at MAMI. These new cross section data 
have almost doubled the world $\gamma n\rightarrow\pi^0n$ database below 
$E$ = 2.7~GeV~\cite{SAID}.

The organization of this paper is as follows. In Sec.~\ref{sec:setup},
details of the A2 experiment and the A2 detector are given. 
Section~\ref{sec:data} outlines the event selection and Sec.~\ref{sec:fsi} 
reviews the approach for determining the final state interaction 
corrections. Section~\ref{sec:res} presents and discusses the measured 
differential cross sections for the reaction $\gamma n\rightarrow\pi^0n$.
Section~\ref{sec:pwa} and L+P fit of the multipoles with determination 
of pole positions and residues. Finally, Sec.~\ref{sec:sum} provides a 
summary of this work and the conclusions.

\section{Experimental Setup}
\label{sec:setup}

\begin{figure}
\begin{center}
\includegraphics[height=3in, keepaspectratio]{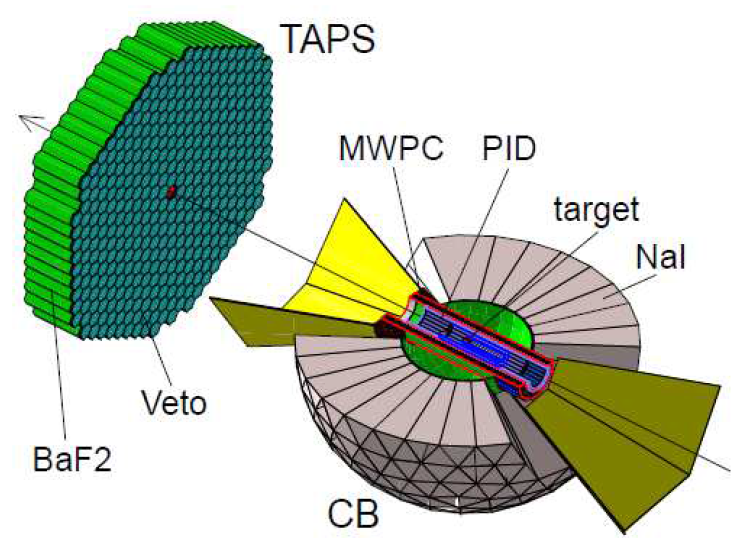}
\end{center}

\vspace{-5mm}
\caption{Set-up of the A2 experiment. CB is a NaI(Tl) calorimeter,
	TAPS is a BaF$_2$ calorimeter which not used for the present 
	measurement, PID is a plastic scintillator detector for 
	particle identification, the MWPC are two cylindrical 
	multiwire proportional chambers, target is a liquid deuterium 
	target (see text for details).} 
	\label{fig:setup}
\end{figure}

The process $\gamma d\rightarrow\pi^0np$ was measured using the Crystal 
Ball (CB)~\cite{Starostin:2001zz} as the central spectrometer
(Fig.~\ref{fig:setup}). Our study shown that there is a marginal 
contribution from TAPS for the reaction $\gamma n\to\pi^0n$ below 
800~MeV. For that reason, we did not use TAPS in our analysis.
The CB was installed in the tagged bremsstrahlung photon beam of the 
Mainz Microtron (MAMI)~\cite{Herminghaus:1983nv,Kaiser:2008zza}, with 
the photon energies determined by the Glasgow tagging 
spectrometer~\cite{Anthony:1991eq,Hall:1996gh,McGeorge:2007tg}.
\vspace{10mm}

The CB spectrometer is a sphere consisting of 672 optically isolated 
NaI(Tl) crystals, shaped as truncated triangular pyramids, 
which point toward the center of the sphere. Each crystal is 41~cm 
long, which corresponds to 15.7 radiation lengths. The crystals are 
arranged in two hemispheres that cover 93\% of $4\pi$~sr, sitting 
outside a central spherical cavity with a radius of 25~cm, which is 
designed to hold the target and inner detectors. 
The CB calorimeter covers polar angles from 20$^\circ$ to 160$^\circ$ 
with full azimuthal coverage. 
The energy resolution for EM showers in the CB can be described as 
$\Delta E/E = 0.020/(E[GeV])^{0.36}$~\cite{Starostin:2001zz}. Shower 
directions are determined with 
a resolution in $\theta$, the polar angle with respect to the beam 
axis, of $\sigma_\theta = 2^\circ - 3^\circ$, under the assumption 
that the photons are produced in the center of the CB. The 
resolution in the azimuthal angle $\phi$ is 
$\sigma_\theta/\sin\theta$. That is an intrinsic CB resolution, 
while in the experiment, angular resolution in $\theta$ is mainly 
defined by the target length.
The CB calorimeter is well suited for detection of both charged 
particles and $\gamma$-quanta. Simultaneously, CB can be used to 
detect neutrons in a wide range of energies~\cite{Stanislaus:2001xk,
Bulychjov:2018mar}.

The Mainz Microtron, MAMI, is a four-stage accelerator, and its
latest addition (the fourth stage) is a harmonic double-sided electron
accelerator~\cite{Kaiser:2008zza}. In this experiment, only the first 
three accelerator stages were used to produce an 883~MeV electron 
beam. Bremsstrahlung photons, 
produced by electrons in a 10-$\mu$m Cu radiator and collimated by a
4-mm-diameter Pb collimator, were incident on a 10-cm-long and
4-cm-diameter liquid deuterium target (LD$_2$) located in the center
of the CB. The energies of the incident photons were analyzed up to
813~MeV by detecting the post-bremsstrahlung electrons in the Glasgow
Tagger~\cite{Anthony:1991eq,Hall:1996gh,McGeorge:2007tg}.

The Tagger is a broad-momentum-band, magnetic-dipole spectrometer that 
focuses post-bremsstrahlung electrons onto a focal-plane detector, 
consisting of 352 half-overlapping plastic scintillators. 
The energy resolution of the tagged photons, which is about $\pm$1~MeV, 
is largely defined by the overlap region of two adjacent scintillation 
counters (a tagger channel) and the electron-beam energy 
used~\cite{McGeorge:2007tg}. For a beam energy of 883~MeV, a tagger 
channel has a width of about 2~MeV for a photon energy 
707~MeV~\cite{McGeorge:2007tg}. Tagged photons are selected in the 
analysis by examining the correlation in time between a tagger channel 
and the experimental trigger derived from CB signals.

The LD$_2$ target is surrounded by a particle identification
detector (PID)~\cite{Watts:2005ei}, which is a cylinder of length 
50~cm and diameter 12~cm, built from 24 identical plastic 
scintillator segments, of thickness 0.4~cm. Outside the PID, there 
are two multi-wire proportional chambers (MWPC), which measure the 
three-dimensional coordinates of a charged track.  

The experimental trigger had one main requirement - the sum of the 
pulse amplitudes from the CB crystals had to exceed a hardware 
threshold that corresponded to an energy deposit larger than 40~MeV.
To select the reaction $\gamma n\rightarrow\pi^0n$, we require that 
the final $\pi^0$ and neutron were detected by the CB.  In this case, 
the number of clusters which fire, i.e., groups of adjacent crystals 
in which energy is deposited by a particle's interaction with the 
calorimeter, is equal to three. 
\begin{figure}
\begin{center}
\includegraphics[height=4in, keepaspectratio, angle=90]{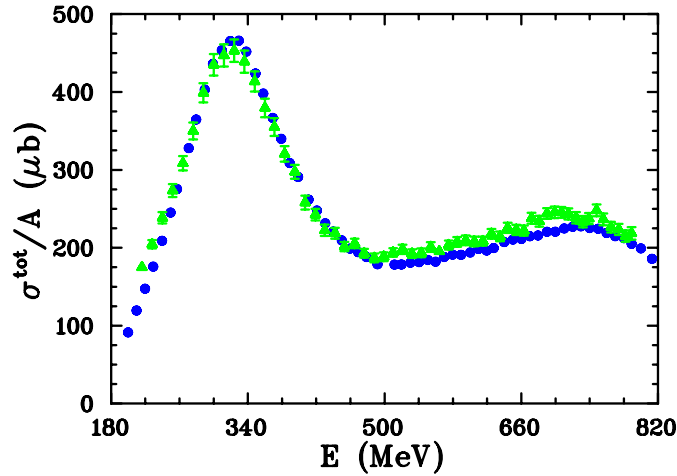}
\end{center}

\vspace{-10mm}
\caption{The total photoabsorption cross section per nucleon
	as a function of incident photon energy. 
	Filled blue circles (filled green triangles) are new A2 
	(previous DAPHNE~\protect\cite{MacCormick:1996jz,
	MacCormick:1997ek}) measurements. Only statistical 
	uncertainties are shown for all the data.
	The total normalization uncertainties of our data are
	about 3\%.} \label{fig:phab}
\end{figure}


In order to provide a check on the performance of the CB detector, 
and to evaluate the efficiency of the trigger, a comparison of the 
total measured cross section, after the empty target background was 
subtracted, was made with the previously published total 
photoabsorption cross section measured by the DAPHNE experiment at 
MAMI~\cite{MacCormick:1996jz,MacCormick:1997ek}. This comparison is 
shown in Fig.~\ref{fig:phab} for photon energies from 180~MeV to 820~MeV. 
In the $\rm\Delta(1232)$ region, the new and previous data are in 
agreement within the systematic uncertainties of the DAPHNE experiment 
(2.5\% -- 3.0\%) and the new measurements ($\sim$6\%). Above 500~MeV, 
our cross section data fall slightly below the previous measurements. 
The difference reaches 6\% at 700~MeV. We do not apply any correction 
for the acceptance of the CB detector, but both the CB and DAPHNE 
detectors do have a similar inefficient region at forward angles. The 
correction for this inefficiency has been extensively studied in 
Ref.~\cite{Watts:2005ei}, where it is shown that the dominant 
contribution comes from single charged pion production, and the value 
of this correction is $\sim$6\%. Taking this into account, our data are 
in satisfactory agreement with the DAPHNE total photon absorption cross 
sections over the full photon energy range. This result demonstrates a 
high efficiency of the Crystal Ball detector for registering secondary 
particles and that the background present in the current measurement 
is small.

\section{Data Handling}
\label{sec:data}

The collected data allowed a detailed study of the reaction 
dynamics.  The $\gamma n\rightarrow\pi^0n$ differential cross 
sections were determined for 27 energy bins and the full range 
of production angles using model-dependent nuclear corrections 
to determine $\gamma n\rightarrow\pi^0n$ data from measurements on 
the liquid deuterium target. 

The main source of the background for our neutral channel is the 
reaction $\gamma p\rightarrow\pi^0p$, where the high-energy proton 
was recorded by the PID detector, and such events were discarded 
from further analysis. Then the selection of the neutral pion is 
based on the search for two photons with an effective mass 
$m_{\gamma\gamma}$, close to the nominal mass of $\pi^0$ in the 
mass interval of 50 to 200~MeV.  Reconstructed neutral pion and 
additional clusters in the CB calorimeter, to determine the 
direction of neutron emission, give complete reaction kinematics 
for $\gamma d\rightarrow\pi^0np$.

\begin{figure}
\begin{center}
\includegraphics[height=3.5in, keepaspectratio]{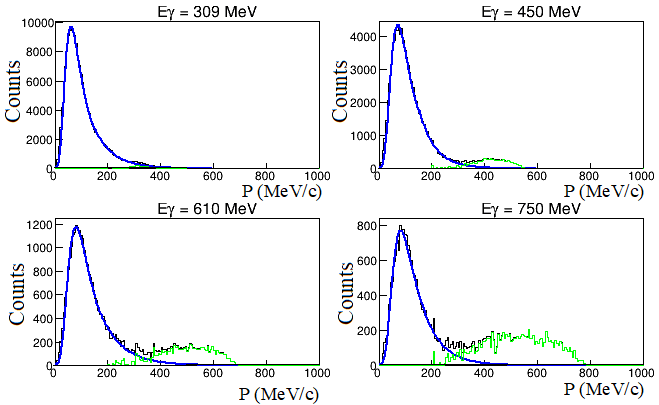} 
\end{center}

\vspace{-5mm}
\caption{Momentum distribution for low-energy protons 
	for four initial photon energies.  Experimental 
	distributions (black) were fitted by a Landau-function 
	in the peak region and an exponential to the right of 
	the peak (blue). The background contribution is shown
	in green.} \label{fig:mom}
\end{figure}

Analysis of the shape of the proton momentum spectrum allows us to 
determine the background contribution arising rom the production 
of two or more neutral pions. The distribution of the proton 
momentum spectrum is shown in Fig.~\ref{fig:mom}, where the low 
momentum component, corresponding to the Fermi-momentum of 
nucleons in the deuteron~\cite{Cool:1970fn}, can be seen. It can be 
well described by the superposition of the 
Landau-function~\cite{Landau:1944if} in the peak region and an 
exponential to the right of the peak. Thus, the background can be 
removed for the component located in the high-momentum region on 
the right side of the distribution. The background significantly 
increases with the initial photon energy and reaches up to 
$\sim$40\% at 800~MeV while it is almost zero at 200~MeV.
\vspace{10mm}

\begin{figure}
\begin{center}
\includegraphics[height=5cm, keepaspectratio]{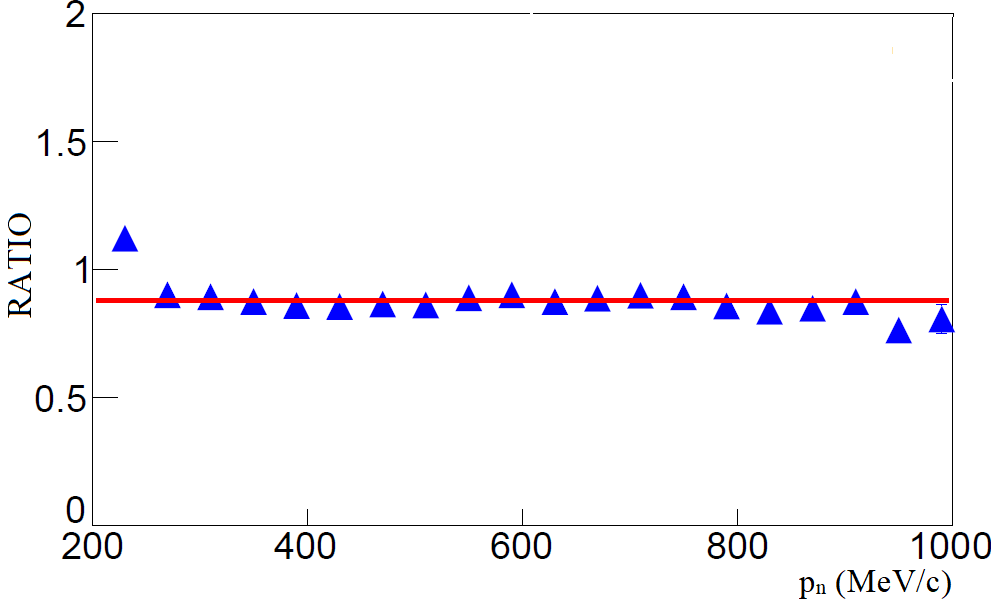}
\includegraphics[height=5cm, keepaspectratio]{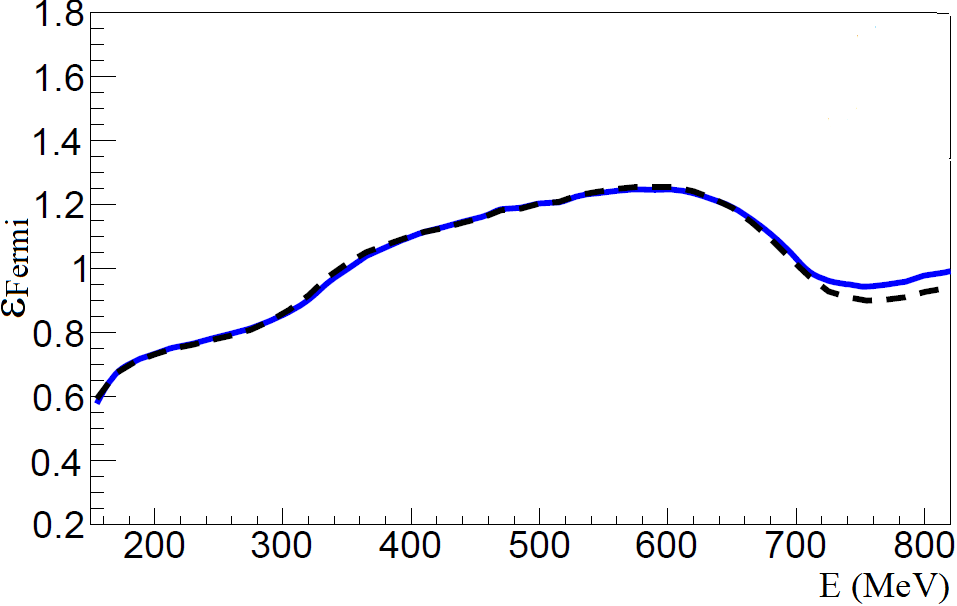} 
\end{center}

\vspace{-5mm}
\caption{Corrections for the reaction $\gamma n\rightarrow\pi^0n$ 
	cross sections. 
	\underline{Left}: Ratio of the efficiency of neutron 
	detection to a simulation of the CB calorimeter vs 
	neutron momentum. Solid red line corresponds to
	the average value of the ratio = 0.878$\pm$0.004.
	\underline{Right}: The Fermi-momentum correction 
	$\varepsilon_{Fermi}$ in the deuteron for recent SAID 
	(blue solid line) and MAID (black dashed line) solutions 
	vs initial photon energy.} \label{fig:neutron}
\end{figure}

\begin{figure}
\begin{center}
\includegraphics[height=5.5cm, keepaspectratio, angle=90]{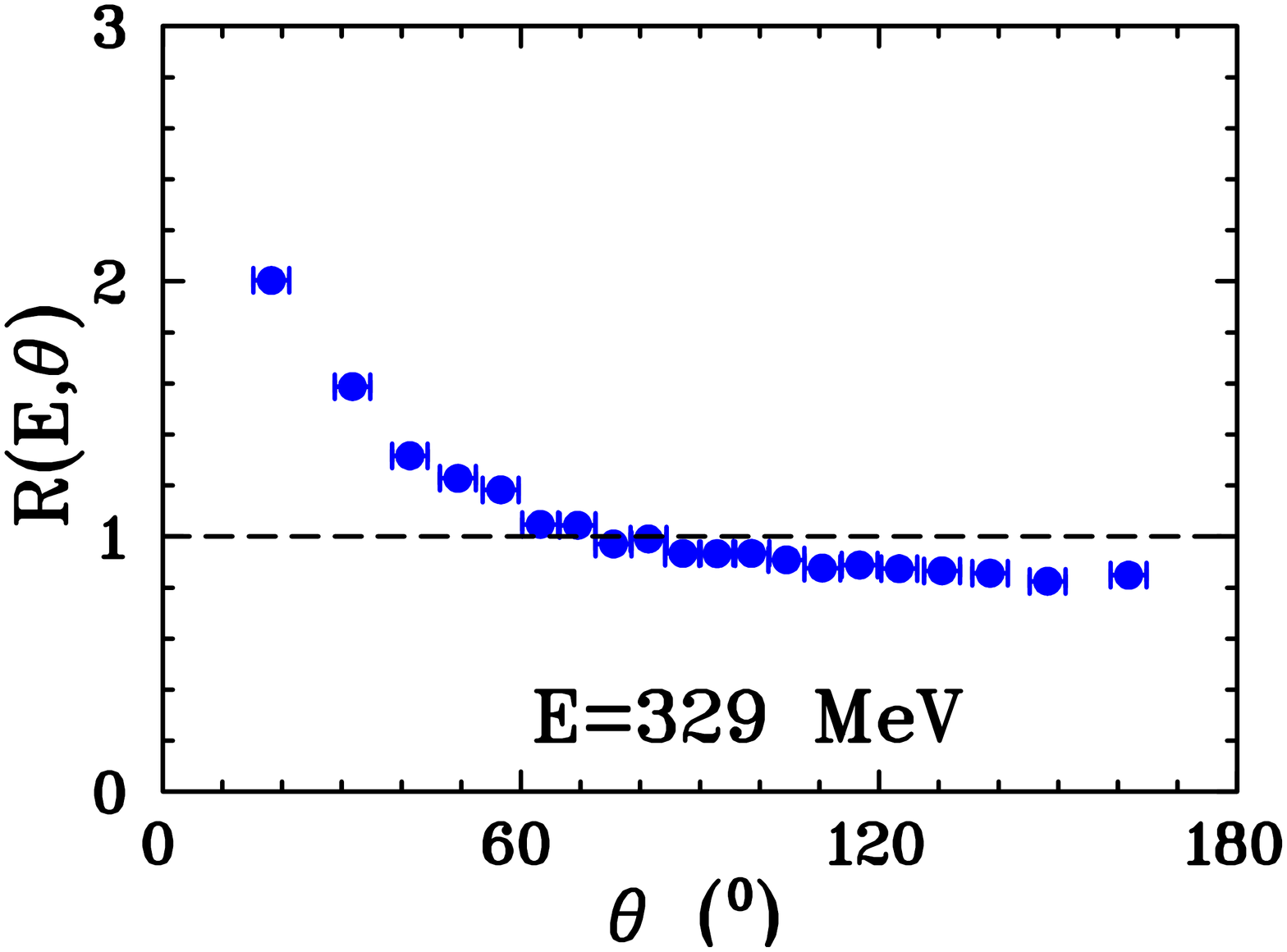}
\includegraphics[height=5.5cm, keepaspectratio, angle=90]{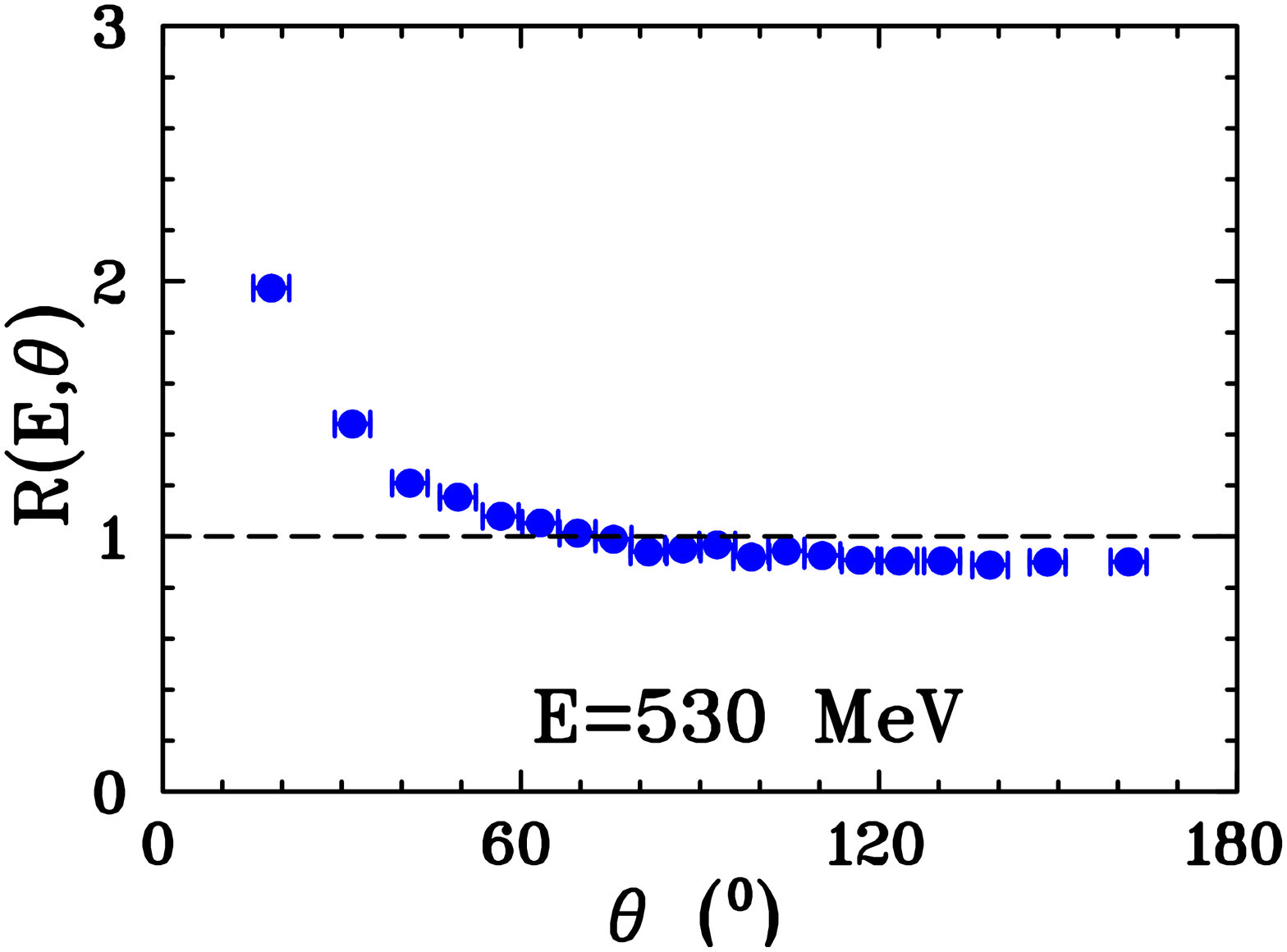}
\includegraphics[height=5.5cm, keepaspectratio, angle=90]{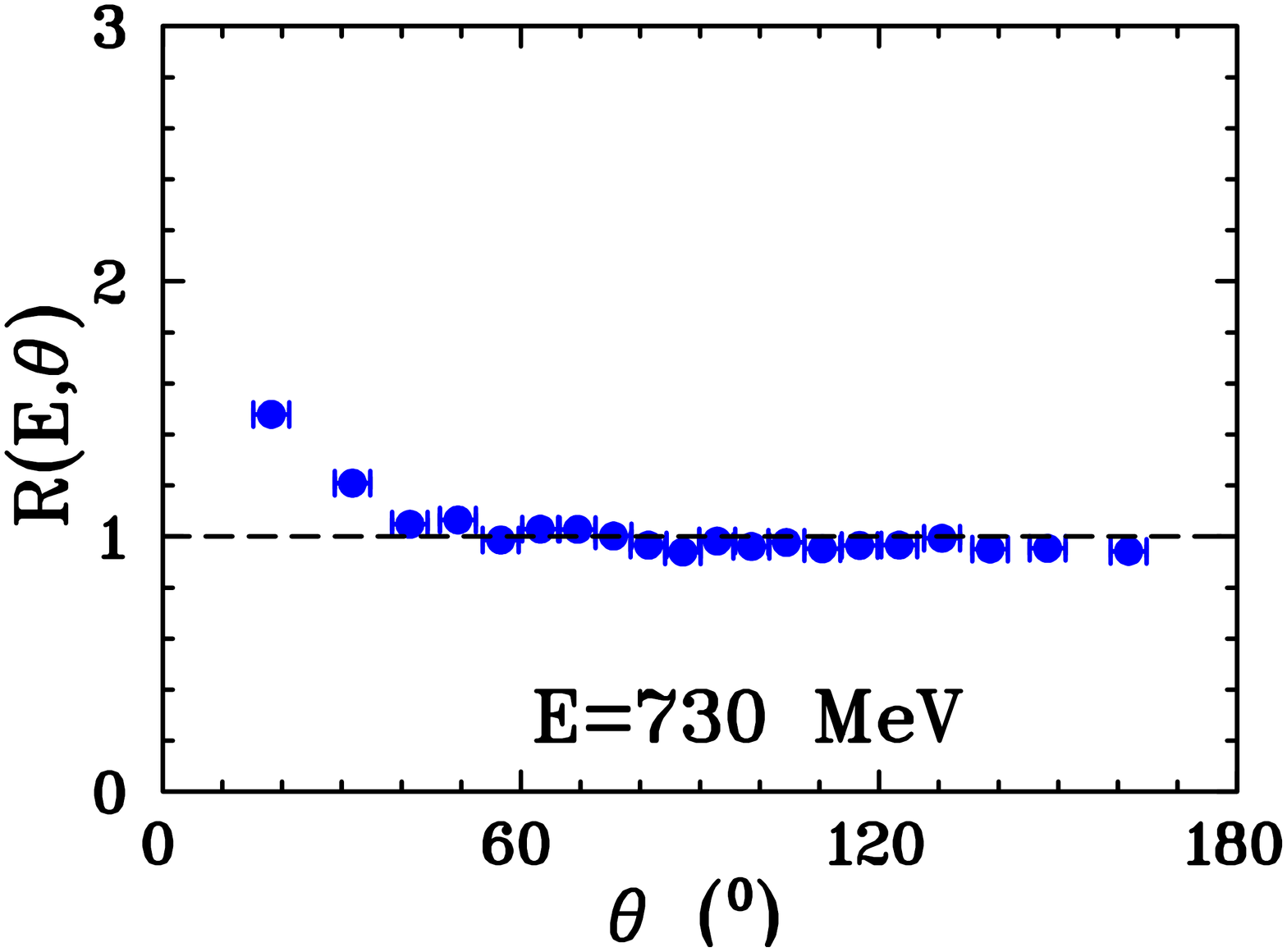}
\end{center}

\vspace{-5mm}
\caption{The FSI correction factor R(E,$\theta$) for selected 
	beam energies vs the polar angle $\theta$ of the outgoing 
	$\pi^0$ in the rest frame of the $\pi^0$ and the fast 
	neutron. 
	The normalization uncertainties are not shown.} 
	\label{fig:fsi1}
\end{figure}

In this work, much attention was paid to the efficiency of 
neutron detection, knowledge of which is necessary to determine 
reaction cross sections. 
The thickness of CB crystals is one interaction length only, so
it is impossible to measure neutron energy, however, it is possible 
to determine the location of the neutron interaction in CB.  This 
is enough for neutron detection efficiency measurements using the 
same data. It has been done in Ref.~\cite{Martemianov:2015oua} by 
searching for the point of neutron interaction in the CB within a 
predicted direction from the kinematics of pion photoproduction on 
the deuteron with known kinematic parameters of a neutral pion and 
proton.  It should be noted that the analysis of data at low photon 
energies is a rather difficult task.  In this case, the neutron 
detection efficiency decreases with decreasing photon and neutron 
energies and is only a few percent at 20 -- 30~MeV neutron 
energies, which can lead to inaccuracy in the analysis of 
experimental data. In addition to the CB calorimeter, the neutron 
detection threshold per cluster affects the efficiency of neutron 
detection. The ratio of the efficiency of neutron detection to 
simulation is shown in Fig.~\ref{fig:neutron} (left) as a function
of neutron momentum.  To determine a full $\gamma 
n\rightarrow\pi^0n$ reaction efficiency, the software package 
Geant4~\cite{Allison:2016lfl} was used in the simulation of 
the experiment.  The event generator used a theoretical model 
based on a detailed description of the deuteron structure, taking 
into account the Fermi motion of nucleons in the deuteron and 
NN-FSI effects. To compare the results for the cross sections with 
the PWA predictions of SAID~\cite{Mattione:2017fxc} and 
MAID~\cite{Drechsel:2007if}, where the cross sections are given 
for the free neutron and neutron at rest 
(Fig.~\ref{fig:neutron} (right)), it is necessary to take into 
account the corresponding corrections~\cite{Bulychev:2018yjy}.  
The calculations used corrections determined from the MAID2007 
analysis. Corrections for the interaction effect in the 
final-state depending on the $\pi^0$ production angle in c.m.\ for 
different values of the incident photon energy are presented in 
Fig.~\ref{fig:fsi1}. The effect of the FSI correction becomes 
significant especially, at low photon energies. 
\vspace{10mm}

The photon flux is defined by counting the scattered electrons 
in the focal plane of the tagging system and correcting for the 
emitted photons lost in the collimator. The probability for 
bremsstrahlung photons to reach the target is measured 
periodically during data taking by a total-absorption lead glass 
counter, which is moved into the photon beam line at reduced 
photon intensity. Using this method, the tagging efficiency for 
an 883~MeV electron beam was determined to be approximately 35\% 
with an accuracy of about 5\%.

One of the main contributions to systematic uncertainties is the 
definition of the photon flux, which is based on the calculation 
of tagging efficiency. Another source of uncertainty is the 
background subtraction due to the empty target. This contribution 
is not large and has an order of 1\%. The systematic uncertainty 
of the total photoabsorbtion cross section was defined by these 
two factors.

The analysis cuts and Monte Carlo simulations, used to obtain the 
$\gamma n\to\pi^0n$ total cross section, introduced a further 
uncertainty of the order of 1\% -- 2\%. Furthermore, the accuracy 
of the neutron detection efficiency leads to an additional 
uncertainty at low energies of $\sim$3\%. So, the total overall 
uncertainty for the $\gamma n\to\pi^0n$ total cross section is 
about 6\%.

In the case of the $\gamma n\to\pi^0n$ differential cross sections, 
the largest systematic uncertainties appear at forward and backward 
pion angles, where the statistical errors increase. This effect is 
actually due to the low efficiency of neutron detection. The typical 
systematic uncertaintes at these forward and backward pion angles 
are estimated to be $\sim$10\%.


\section{Final-State Interaction}
\label{sec:fsi}

\begin{figure}
\begin{center}
\includegraphics[height=3cm, keepaspectratio]{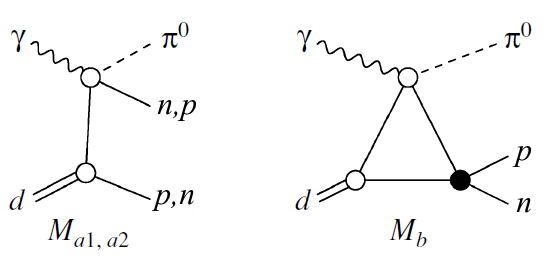}
\end{center}

\vspace{-5mm}
\caption{Feynman diagrams for the leading terms of the $\gamma
        d\rightarrow\pi^0np$ amplitude: the impulse 
	approximation ($M_{a1}, M_{a2}$) and $NN$-FSI ($M_b$). 
	The wavy, dashed, solid, and double lines correspond 
	to the photons, pions, nucleons, and deuterons, 
	respectively.} \label{fig:fsi}
\end{figure}

Exact determinations of the differential pion photoproduction cross 
sections on the neutron, based on experimental data for $\gamma 
d\rightarrow\pi^0pn$, can not be implemented without a reliable 
theoretical reaction model. This model was developed, taking into 
account the contribution not only of IA (Fig.~\ref{fig:fsi} (left)), 
but also effects of $NN$ and $\pi N$ interactions in the final 
state. A detailed description of FSI effects is given in 
Refs.~\cite{Tarasov:2011ec,Tarasov:2015sta}. The SAID 
phenomenological amplitudes, from $NN$~\cite{Arndt:2007qn} and $\pi 
N$~\cite{Arndt:2006bf} elastic with $\gamma N\rightarrow\pi 
N$~\cite{Dugger:2007bt} PWAs, were used as input for the GWU-ITEP 
code. The full Bonn potential~\cite{Machleidt:2000ge}  was used 
for the deuteron description and Fermi motion of nucleons in the 
deuteron was taken into account. In this paper, to speed up 
numerical calculations the model has been simplified. The 
contribution from $\pi N$-FSI, which is important close to the 
threshold, does not play a significant role above $E$ = 200~MeV, 
as claimed in Ref.~\cite{Levchuk:2006vm}, and so was neglected 
here.  The effect of NN-FSI was taken into account in the 
S-wave approximation which makes the dominant contribution. The 
parameters of the $pn$-scattering $s$-wave amplitudes with 
isospins $0$ and $1$ were taken from Ref.~\cite{Landau:1991wop}. 
Thus, the amplitude $M$ of the reaction $\gamma d\rightarrow
\pi^0pn$ (Fig.~\ref{fig:fsi}) is given as
\begin{eqnarray}
	M = M_{a1} + \rm\Delta,~~~ \rm\Delta = M_{a2} + M_b,
        \label{eq:fsi1}
\end{eqnarray}\noindent
where $M_{a1}$ is a leading IA diagram with the fast neutron and 
$\rm\Delta$ is a correction that takes into account the 
$pn$-FSI ($M_b$) and IA diagram with regrouped nucleons ($M_{a2}$). 
The expressions for these amplitudes are given in detail in 
Appendix~A of Ref.~\cite{Tarasov:2015sta} (Note that we did not 
include the off-shell factor for the $\gamma N\rightarrow\pi N$ 
amplitude, introduced in Ref.~\cite{Tarasov:2015sta}, Eq.~(18)).  
Calculating such a correction for the analysis of the experimental 
data means taking into account each event with a weight as
\begin{eqnarray}
	R = \frac{\overline{|M_{a1}|^2}}{\overline{|M|^2}},
        \label{eq:fsi2}
\end{eqnarray}\noindent
where $\overline{|M_{a1}|^2}$ and $\overline{|M|^2}$ are amplitude 
squares averaged over spins and calculated for the kinematics of 
the events. Furthermore, event handling was carried out under the 
assumption that the reaction mechanism is determined by the 
diagram $M_{a1}$.
\vspace{10mm}
\vspace{10mm}

So, we will determine the differential cross sections of the 
reaction $\gamma n\rightarrow\pi^0n$ from the measurements on the 
deuteron using the theoretical model, which describes the reaction 
$\gamma d\rightarrow\pi^0pn$.  

\section{Experimental Results}
\label{sec:res}

The differential cross section for the $\gamma n\rightarrow\pi^0n$
reaction is defined by the formula:
\begin{equation}
	\frac{d\sigma(E, \theta)}{d\Omega}=\sigma_{0}\frac{\rm\Delta 
	N_{Events}}{N_{\rm{Scaler}}~\varepsilon_{\rm{Tag}}}
	\times\frac{1}{\varepsilon_{\rm{Sel}}~\varepsilon_{\rm{Backg}}
	~\varepsilon_{\rm{PID}}~\varepsilon_{\rm{Fermi}}
	~\varepsilon_{\rm{FSI}}~2\pi~dcos(\theta)}
	\label{eq:dsg}
\end{equation}
where
$\sigma_{0}$~=~$\rho_{\rm{targ}}L_{\rm{targ}}N_{A}/2\times10^{-7}$~($\mu b$)
        is the nuclear density of the liquid deuterium target;
$\rm\Delta N_{Events}$ is the number of the events in $d\cos(\theta)$,
	where $\theta$ is the neutral pion production angle in c.m.\
	relative to the beam axis;
$N_{\rm{Scaler}}$ is the number of counts in the tagger scalers;
	$\varepsilon_{\rm{Tag}}$ is the tagging efficiency (fraction of 
	photons impinging on the target);
$\varepsilon_{\rm{Sel}}$ is the selection efficiency obtained from the 
	simulated data, which also includes neutron detection efficiency 
	incorporated into the Monte-Carlo.  
	The neutron detection efficiency is defined directly from data 
	by the method proposed in Ref.~\cite{Martemianov:2015oua};
$\varepsilon_{\rm{Backg}}$ is the background extraction efficiency, which 
	included the empty target correction and the correction for random 
	photons in the beam;
$\varepsilon_{\rm{PID}}$ is the correction factor for the PID inefficiency, 
	this value was estimated on the experimental data from CB and MWPC 
	and required to select a neutral decay channel;
$\varepsilon_{\rm{FSI}}$ is the FSI-effect correction to get the $\gamma 
	n\rightarrow\pi^{0}n$ cross section from the deuteron
	measurements, $\varepsilon_{\rm{FSI}} = R(E,\theta)^{-1}$
	(see above).
\begin{figure}
\begin{center}
\includegraphics[height=7cm, keepaspectratio, angle=90]{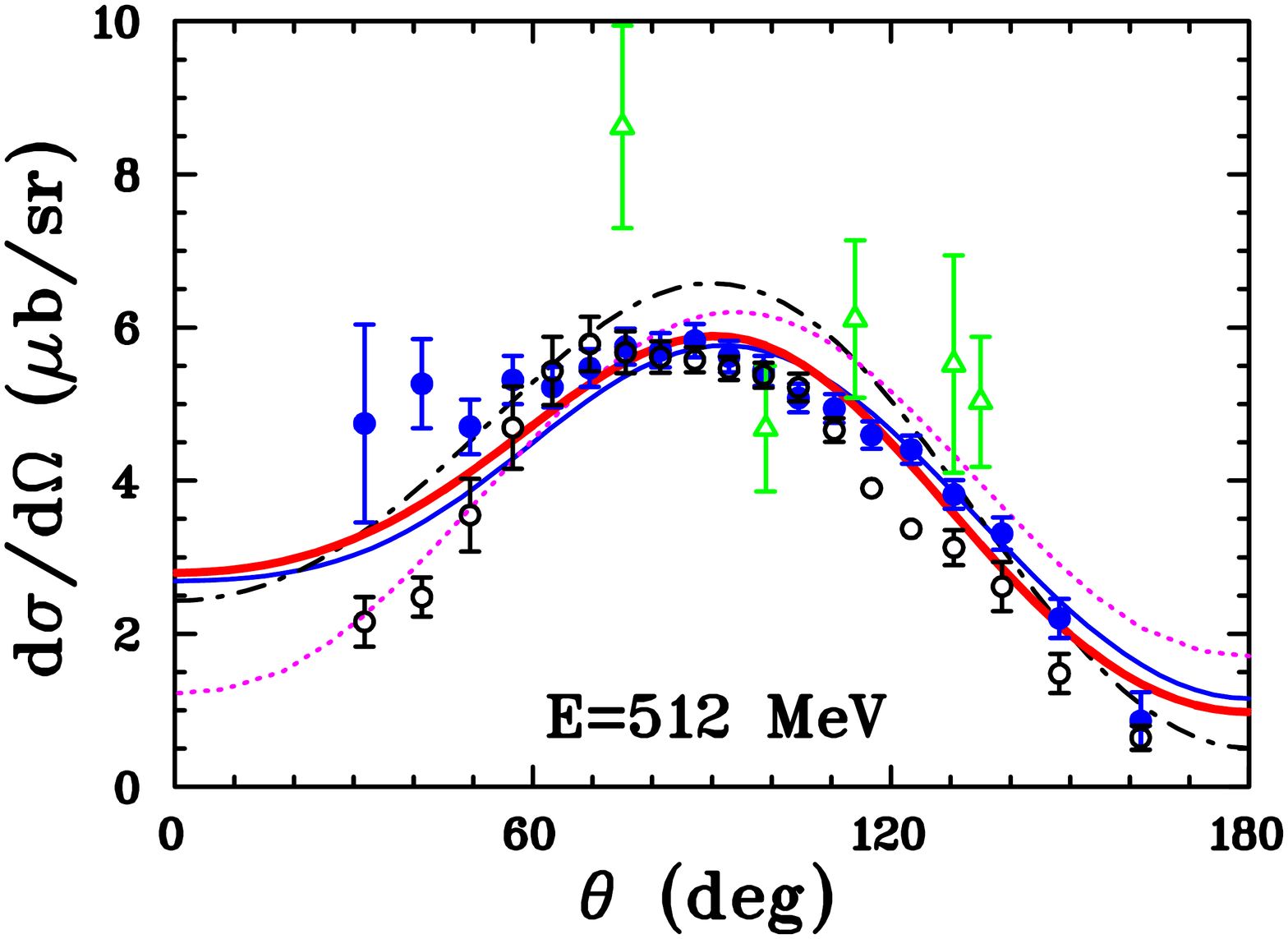}
\includegraphics[height=7cm, keepaspectratio, angle=90]{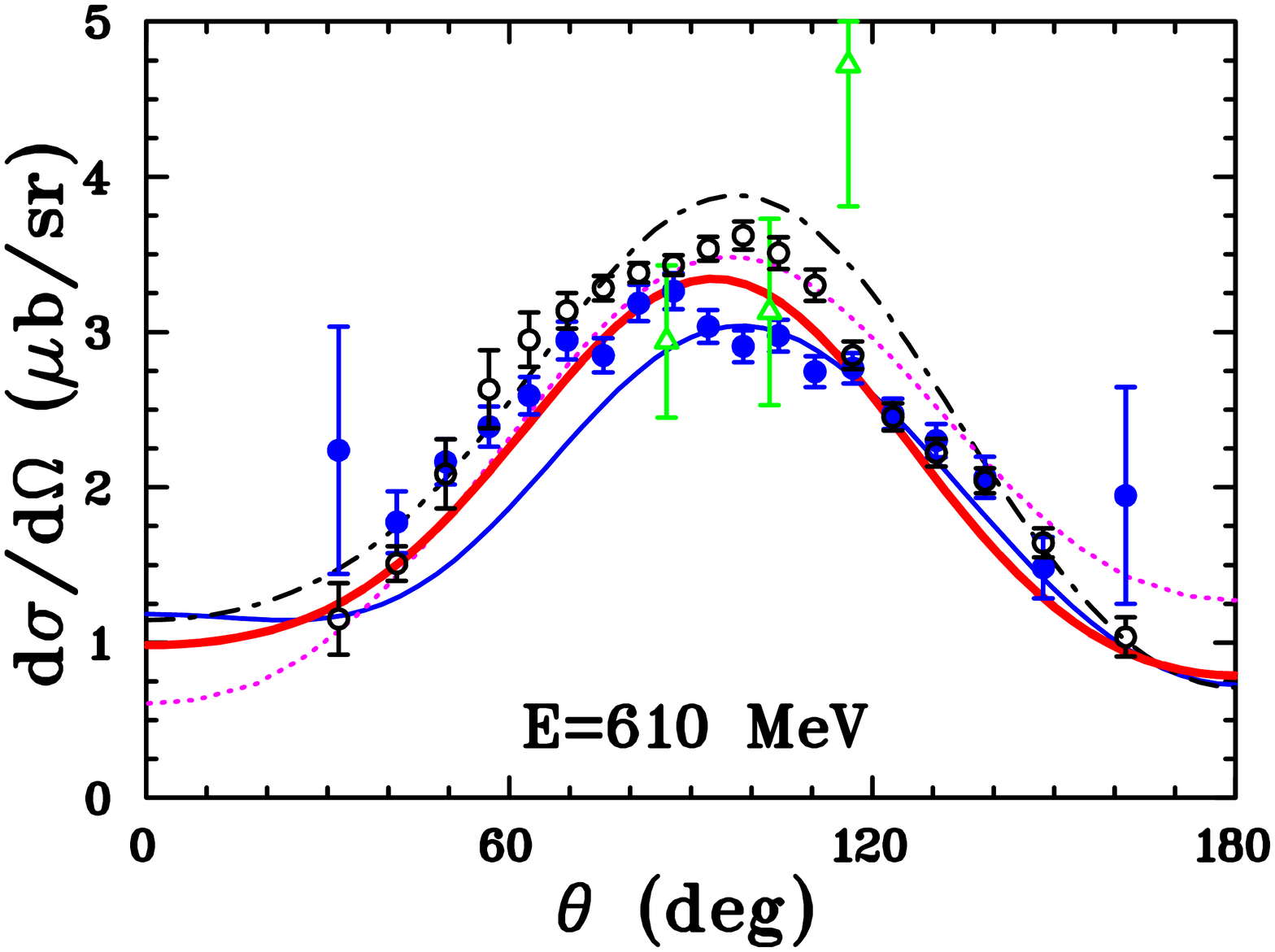}
\includegraphics[height=7cm, keepaspectratio, angle=90]{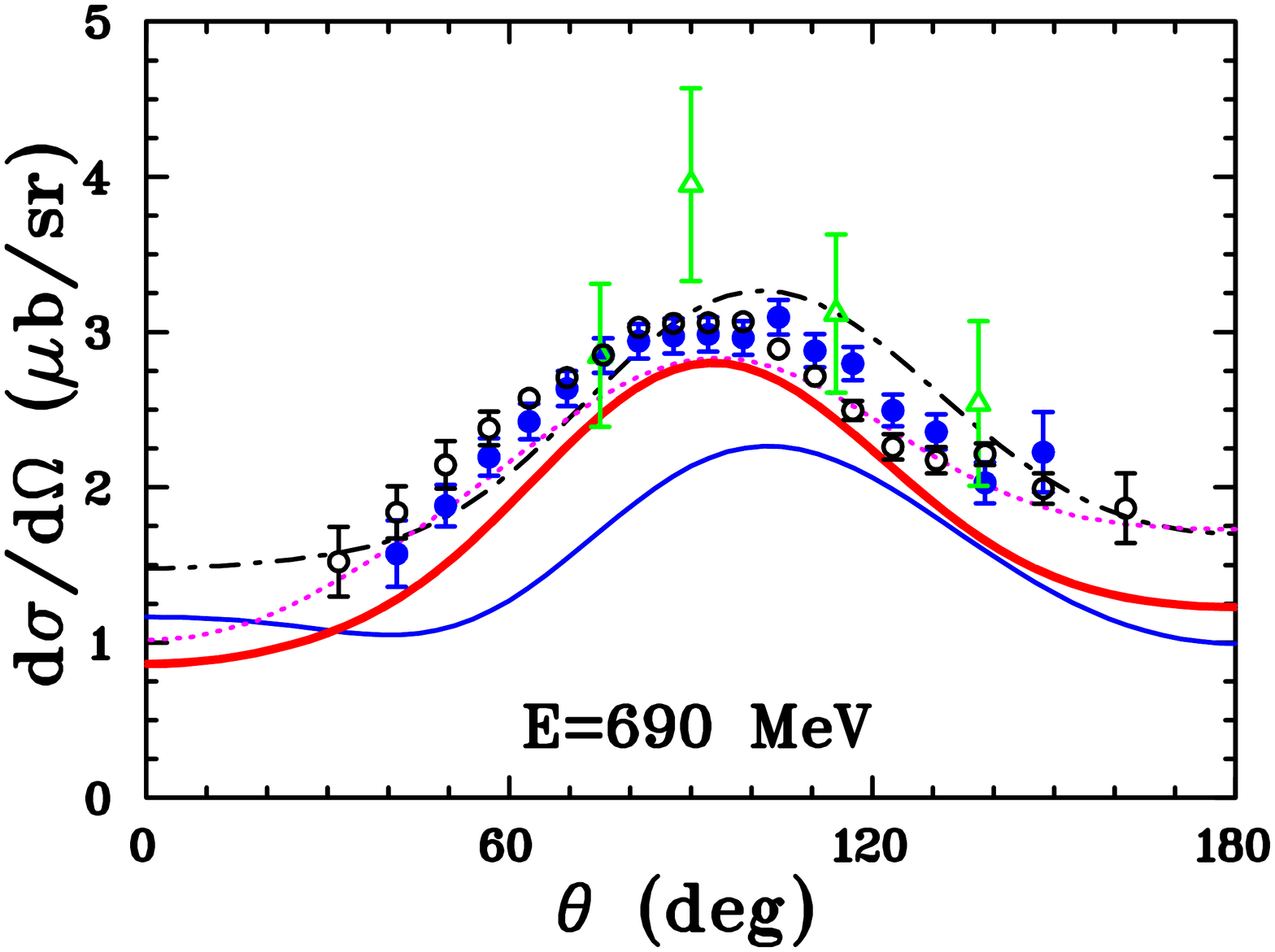}
\includegraphics[height=7cm, keepaspectratio, angle=90]{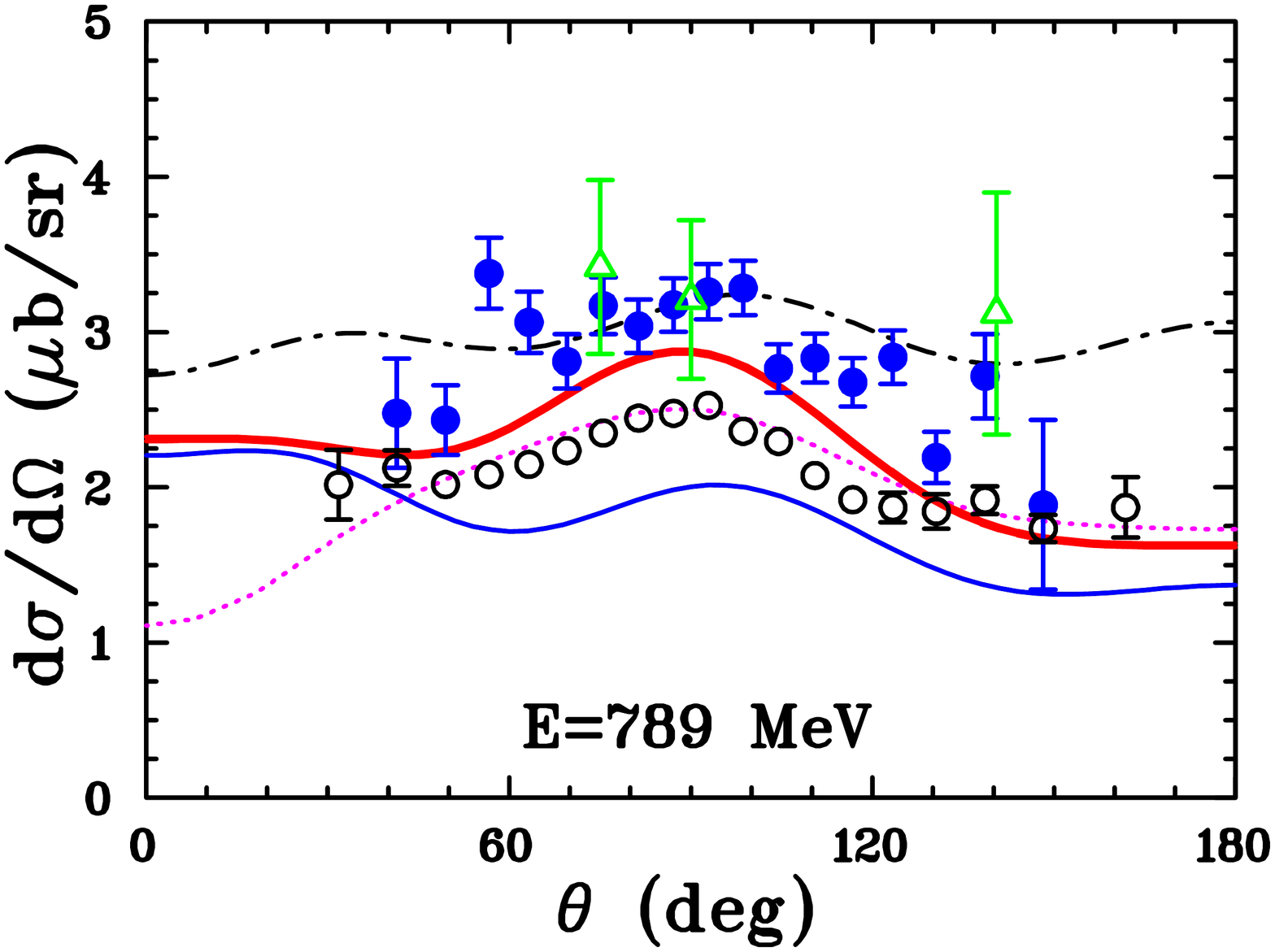} 
\end{center}

\vspace{-7mm}
\caption{Differential cross section for the reaction 
	$\gamma n\rightarrow\pi^0n$ vs $\pi^0$ c.m.\
	polar angle for selectedphoton energies.
        \underline{Data}: filled blue (open black) circles are
        new (previous~\protect\cite{Dieterle:2018adj}) 
	A2 measurements, open green triangles are the
        previous non-MAMI measurements~\protect\cite{SAID}.
        Only angle-dependent (statistical and systematical 
	combined in quadratures) uncertainties are shown for 
	all the data. The total normalization (angular 
	independent) uncertainties of the cross section vary 
	between 2 and 6.5\% (not included in plots).
        \underline{Fits}: SAID MA19 (red thick solid lines),     
        recent SAID MA27 (blue solid
        lines)~\protect\cite{Mattione:2017fxc}, MAID2007
        (black dash-dotted lines)~\protect\cite{Drechsel:2007if},
        and BnGa BG2014-02 (magenta dotted
        lines)~\protect\cite{Werthmuller:2014thb}.}
        \label{fig:dsg0}
\end{figure}


Since our results for the $\gamma n\rightarrow\pi^0n$ differential 
cross sections consist of 492 experimental points, they are not
tabulated in this publication but are available in the SAID
database~\cite{SAID} along with their uncertainties and the energy
binning. 

In Fig.~\ref{fig:dsg0}, our differential cross sections for four 
incident photon energies are compared with previous measurements
made at similar energies~\cite{SAID}. Some of these 
measurements~\cite{Dieterle:2018adj} are quite 
recent while most of them were performed in the 1970s at Tokyo and 
Frascati bremsstrahlung facilities, demonstrating the general 
desire of the resonance-physics community to obtain new $\gamma 
N\rightarrow\pi N$ data, which are needed for a better determination 
of the properties of the $N^\ast$ states.  
As seen in Fig.~\ref{fig:dsg0}, all our results are in reasonable 
agreement with the previous measurements and significantly extend 
down to lower energies the previously published A2 
data~\cite{Dieterle:2018adj}, covering the $\rm\Delta$-isobar 
resonance peak region.
Some discrepancies are observed between the data 
obtained at forward production angles, but this can be explained 
by the difference in the energy binning of the data sets, bearing 
in mind the rapidly falling cross section close to the forward 
direction where the FSI correction increases (Fig.~\ref{fig:fsi1}).
\begin{figure}[htb!]
\begin{center}
    \includegraphics[height=4in, keepaspectratio, angle=90]{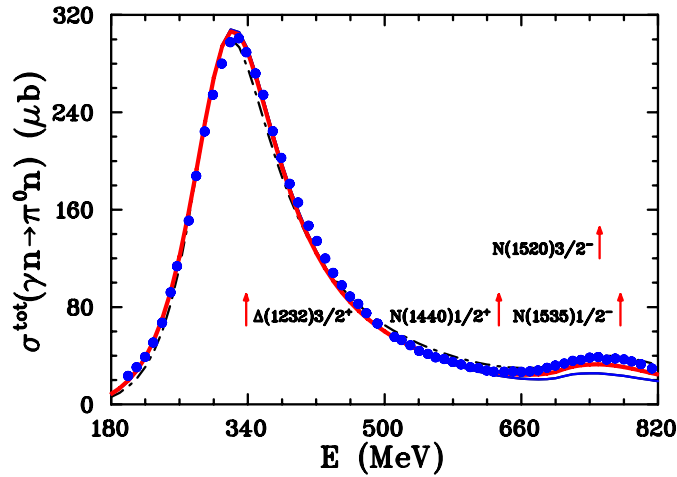}
\end{center}

\vspace{-10mm}
\caption{Total cross section of the reaction
        $\gamma n\rightarrow\pi^0n$ as a function of 
	incident photon energy.  Only statistical
        uncertainties are shown for the data.
        The total normalization uncertainties 
	vary between 2 and 6.5\%.  Vertical red arrows 
	show BW masses of low-lying 
	resonances~\protect\cite{Tanabashi:2018oca}.
        Notation as in Fig.~\protect\ref{fig:dsg0}.}
        \label{fig:sgt}
\end{figure}
\begin{figure*}[htb!]
\begin{center}
    \includegraphics[height=7in, keepaspectratio, angle=90]{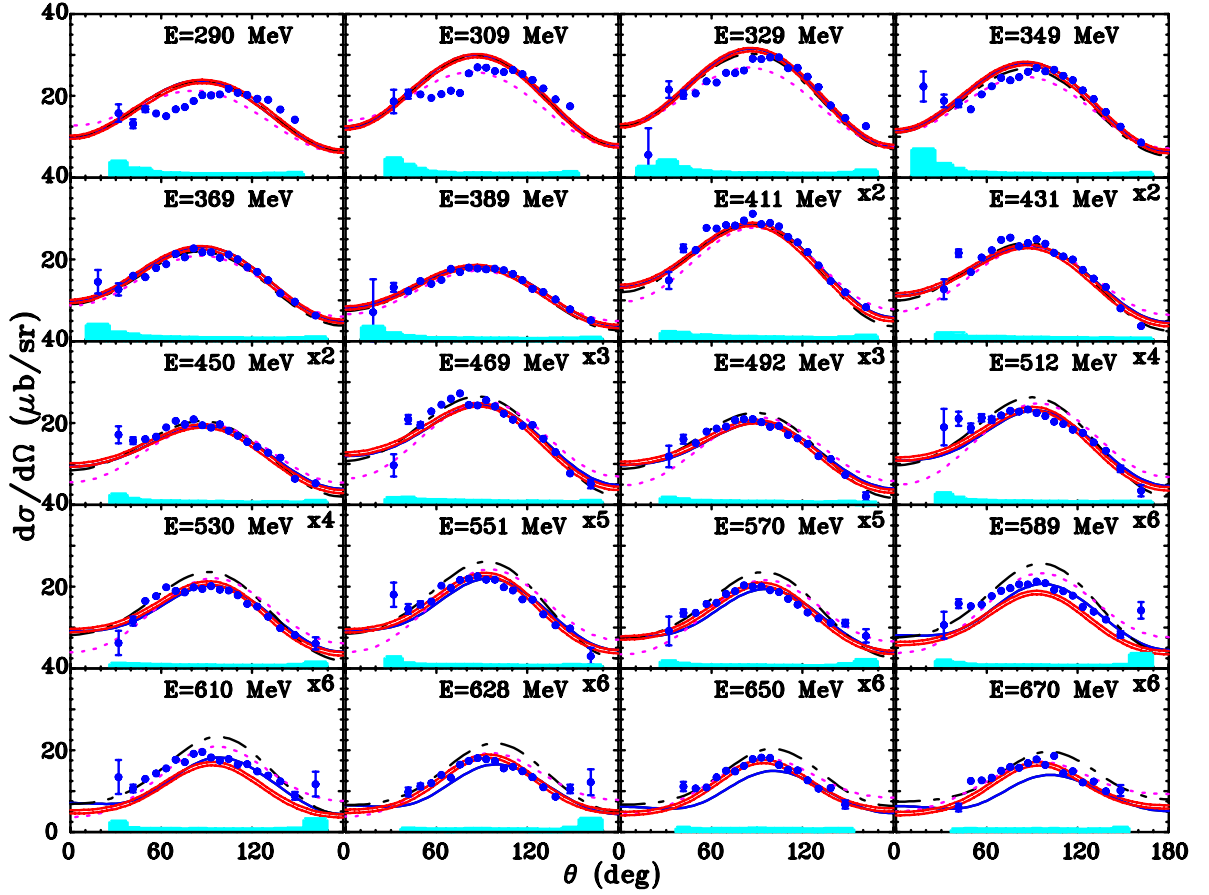}
\end{center}

\vspace{-12mm}
\caption{Differential cross section for the reaction $\gamma 
	n\rightarrow\pi^0n$ below 680~MeV. Notation as in 
	Fig.~\protect\ref{fig:dsg0}. Only statistical 
	uncertainties are shown for all data. The 
	angular-dependent systematic uncertainties are shown 
	by cyan histograms.} \label{fig:dsg1}
\end{figure*}
\begin{figure*}[htb!]
\begin{center}
    \includegraphics[height=7in, keepaspectratio, angle=90]{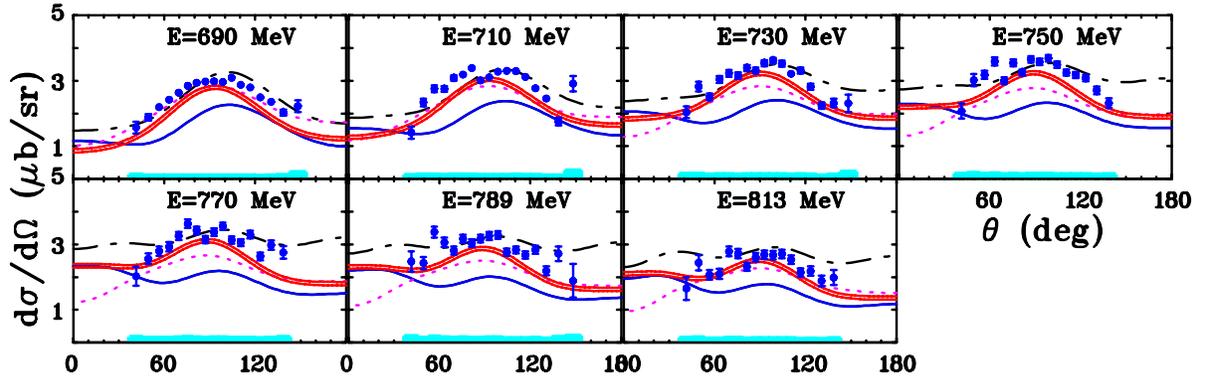}
\end{center}

\vspace{-75mm}
\caption{Differential cross section of the reaction $\gamma 
	n\rightarrow\pi^0n$ above 680~MeV. Notation as in 
	Fig.~\protect\ref{fig:dsg1}.} \label{fig:dsg2}
\end{figure*}
\begin{figure*}[htb!]
\begin{center}
    \includegraphics[height=7in, keepaspectratio, angle=90]{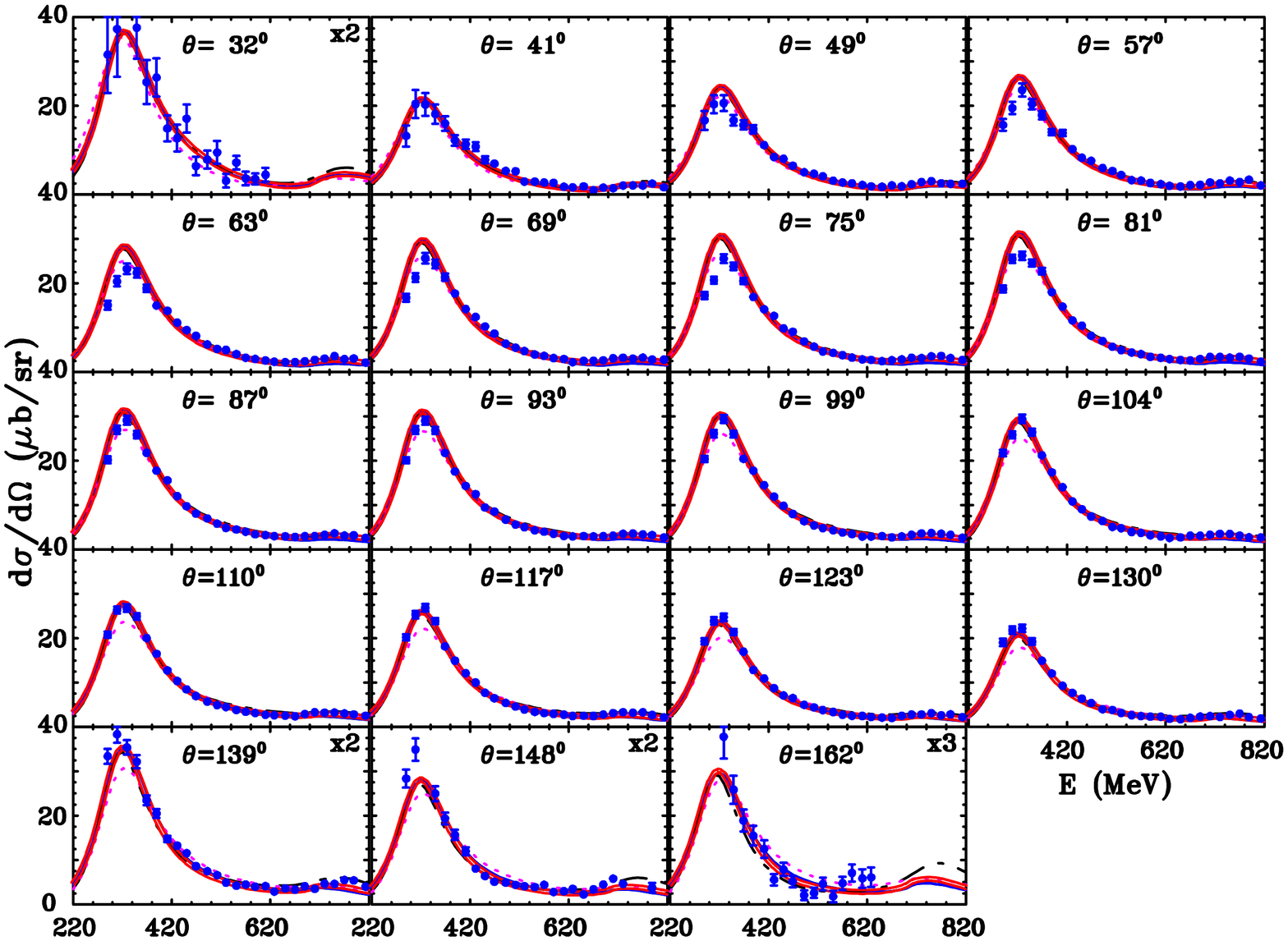}
\end{center}

\vspace{-12mm}
\caption{Fixed-angle excitation function for the 
	$\gamma n\rightarrow\pi^0n$ reaction as a 
	function of incident photon energy. 
        The uncertainties shown are the angle-dependent 
	statistical and systematic uncertainties, 
	combined in quadrature are shown 
	for all the data. Notation as in 
	Fig.~\protect\ref{fig:dsg1}.} \label{fig:dsg3}
\end{figure*}
Presented in Fig.~\ref{fig:sgt} are total cross sections for 
$\gamma n\rightarrow\pi^0n$, compared with PWA fits.  There are 
two structures visible here, the first associated with the
$\rm\Delta(1232)3/2^+$, while a small structure is connected to
the $N(1520)3/2^-$ and $N(1535)1/2^-$.

The new SAID best fit results MA19 are in satisfactory 
agreement with the data at energies exceeding 300~MeV 
(Figs.~\ref{fig:dsg1} - \ref{fig:dsg3}), and have visible 
discrepancies at lower energies (Fig.~\ref{fig:dsg1}).  Two 
reasons for this disagreement can be noted. Firstly, this 
is due to the possible underestimation of the interaction 
effect in the final state, which becomes important in this 
particular energy range. The simplified FSI code, for example, 
does not take into account the contribution of the $\pi N$-FSI 
effects which, as a result, may lead to distortion of the 
shape of the spectra of differential cross sections. Secondly, 
the neutron detection efficiency varies with energy. It 
decreases with decreasing particle momentum and is only a few 
percent (see, for instance, Fig.~9 of 
Ref.~\cite{Martemianov:2015oua}), which can lead to large 
systematic errors.

\section{Impact of the Present Data on Partial-Wave Analysis}
\label{sec:pwa}

The SAID parametrization of the transition amplitude 
$T_{\alpha\beta}$ used in the hadronic fits to the $\pi N$ 
scattering data is given as
\begin{eqnarray}
	T_{\alpha\beta} = \sum_{\sigma}|1 - 
	\overline{K}C|^{-1}_{\alpha\sigma}| \overline{K}_{\sigma\beta},
	\label{eq:Kmat}
\end{eqnarray}\noindent
where $\alpha$, $\beta$, and $\sigma$ are channel indices for the 
$\pi N$, $\pi\rm\Delta$, $\rho N$, and $\eta N$ channels. Here 
$\overline{K}_{\alpha\beta}$ are the Chew-Mandelstam $K$-matrices, 
which are parametrized as polynomials in the scattering energy. 
$C_\alpha$ is the Chew-Mandelstam function, an element of a 
diagonal matrix $C$ in channel space, which is expressed as a 
dispersion integral with an imaginary part equal to the two-body 
phase space~\cite{Arndt:1985vj}.

In Ref.~\cite{Workman:2012jf}, it was shown that this form could be
extended to $T_{\alpha\gamma}$ to include the electromagnetic channel 
as
\begin{eqnarray}
	T_{\alpha\gamma} = \sum_{\sigma}|1 - \overline{K}C|^{-1}_{\alpha\sigma}|
	\overline{K}_{\sigma\gamma} .
        \label{eq:Tmat}
\end{eqnarray}\noindent
Here, the Chew-Mandelstam $K$-matrix elements associated with the 
hadronic channels are kept fixed from the previous SAID solution 
SP06~\cite{Arndt:2006bf}, and only the EM elements are varied. The 
resonance pole and cut structures are also fixed from hadronic 
scattering. This provides a minimal description of the 
photoproduction process, where only the $N^\ast$ and 
$\rm\Delta^\ast$ states present in the SAID $\pi N$ scattering 
amplitudes are included in this multipole analysis.

With each angular distribution, a normalization constant $(X)$ and
its uncertainty $(\epsilon_{X})$ were assigned~\cite{Arndt:2002xv}. 
The quantity $\epsilon_{X}$ is generally associated with the 
systematic uncertainty (if known).  The modified $\chi^{2}$ function 
to be minimized is given by
\begin{eqnarray}
        \chi^{2}=\sum_{i}\left(\frac{{X\theta_{i}-
        \theta_{i}^{{\rm exp}}}}{{\epsilon_{i}}}\right)^{2}
        +\left(\frac{{X-1}}{{\epsilon_{X}}}\right)^{2},
	\label{eq:norm}
\end{eqnarray}\noindent
where the subscript $i$ labels the data points within the 
distribution, $\theta_{i}^{{\rm exp}}$ is an individual measurement, 
$\theta_{i}$ is the corresponding calculated value, and $\epsilon_{i}$ 
represents the total angle-dependent uncertainty. The total $\chi^2$ 
is then found by summing over all measurements. This renormalization 
freedom often significantly reduces the overall $\chi^2$ but may 
over-renormalize cross sections significantly beyond limits indicated 
by the systematic errors.  This effect is evident when comparing the 
MA27 curve to the higher-energy data. By increasing the weight of the 
second term in our modified $\chi^2$ function, this problem was 
avoided. The weight was increased until the fitted renormalization 
factors remained inside limits suggested by the systematic errors. 
For other data analyzed in the fit, such as excitation data, the 
statistical and systematic uncertainties were combined in quadrature 
and no renormalization was allowed.  

\begin{figure*}[htb!]
\begin{center}
    \includegraphics[height=4in, keepaspectratio, angle=90]{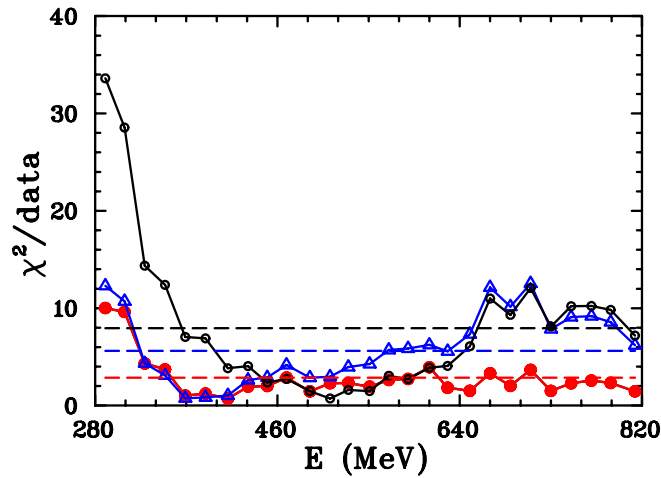}
\end{center}

\vspace{-10mm}
\caption{Comparison of the previous SAID solution 
	MA27~\protect\cite{Mattione:2017fxc} applied to the present 
	A2 data with (blue open triangles) and without (black open 
	circles) FSI corrections and the new SAID MA19 (red 
	full circles) solution obtained after adding the present A2 
	data with FSI corrections into the fit (the solid lines 
	connecting the points are included only to guide the eye). 
	Shown are the fit $\chi^2$ per data point values averaged 
	within each energy bin $E$, where the horizontal dashed lines 
	(blue (black) for MA27 and red for MA19) show the overall 
	$\chi^2$ per data point values.} \label{fig:chi}
\end{figure*}

In the previous fits to the $\gamma n$ differential cross sections, 
the unrestricted best fit gave renormalization constants $X$ 
significantly different from unity. As can be seen from 
Eq.~(\ref{eq:norm}), if an angular distribution contains many 
measurements with small statistical uncertainties, a change in the 
renormalization may improve the fit with only a modest $\chi^2$ 
penalty. Here, however, the weight of the second term in 
Eq.~(\ref{eq:norm}) has been adjusted by the fit for each dataset 
to keep the renormalization constants approximately within 
$\epsilon_{X}$ of unity. This was possible without degrading the 
overall fit $\chi^2$, as can be seen in Fig.~\ref{fig:chi}. 
With the new high-precision $\gamma n\rightarrow\pi^0n$ cross 
sections from the present MAMI dataset, a new SAID multipole analysis 
has been completed. This new global energy-dependent (ED) solution 
has been labeled as MA19. The overall fit quality of the present 
MA19 and previous SAID MA27 solutions are compared in 
Table~\ref{tab:tbl2}.  The inclusion of the present dataset shows 
significant improvement in the comparisons between the SAID MA27 
and MA19 solutions as shown in Fig.~\ref{fig:chi} and 
Table~\ref{tab:tbl2}. This demonstrates the power of these cross 
section measurements with their small uncertainties. The overall 
comparison of the SAID MA19 and MA27 solutions in 
Table~\ref{tab:tbl2} shows that the fit $\chi^2$/data values are 
essentially unchanged for pion photoproduction channels. The 
$\chi^2$/data = 3.77 for MAID2007 vs new present A2 measurements is 
understandable since many recent datasets were not included in this
analysis.
\begin{table}[htb!]

\centering \protect\caption{Average $\chi^2$/data (including 
	information on the total $\chi^2$ from Eq.~(\ref{eq:norm}) 
	and the number of data points used) for 
	MAID2007~\protect\cite{Drechsel:2007if} and two SAID 
	solutions: MA27~\protect\cite{Mattione:2017fxc} and MA19. 
	To satisfy the MAID2007 energy limit, we are presenting 
	results for data below $W$ = 2~GeV or $E$ = 1650~MeV.
	For the $\chi^2$/data of the MAID2007 solution, we took
	into account overall systematic errors as the SAID group
	does~\protect\cite{Arndt:2002xv}.}
\vspace{2mm}
{\begin{tabular}{|c|c|c|c|} \hline
Reaction		    & MAID2007 	         & SAID MA27        & SAID MA19        \tabularnewline
\hline
Present                     & 		         &                  &                  \tabularnewline
$\gamma n\rightarrow\pi^0n$ & 1855/492=3.77      & 2765/492=5.62    & 1405/492=2.86    \tabularnewline
\hline
Previous                    &                    &                  &		       \tabularnewline
$\gamma p\rightarrow\pi^0p$ & 156700/13988=11.20 & 25856/13988=1.85 & 23954/13988=1.71 \tabularnewline
$\gamma p\rightarrow\pi^+n$ & 121150/5225=23.19  & 10785/5225=2.06  & 10371/5225=1.99  \tabularnewline
$\gamma n\rightarrow\pi^-p$ & 49471/4142=11.94   & 7087/4142=1.71   & 6530/4142=1.58   \tabularnewline
$\gamma n\rightarrow\pi^0n$ & 27060/515=52.54    & 2958/515=5.74    & 2320/515=4.51    \tabularnewline
Total			    & 354373/23870=14.85 & 46686/23870=1.96 & 43174/23870=1.81 \tabularnewline
\hline
\end{tabular}} \label{tab:tbl2}
\end{table}

Our next step is to extract the photon decay amplitude at the pole. We do 
this by extracting all the pole positions and residues of the relevant 
partial waves and then we use the residues to obtain the final result, as 
described in Ref.~\cite{Workman:2013rca}. 

Similarly as in Ref.~\cite{Mattione:2017fxc}, the Laurent+Pietarinen (L+P) 
method has been applied to determine the pole positions and residues from the 
pion photoproduction multipoles~\cite{Svarc:2014sqa}. The driving concept 
behind the method is to replace the complexity of solving an elaborate 
theoretical model and analytically continuing its solution into the complex 
energy plane by using a local power-series representation of partial wave 
amplitudes that only exploits analyticity and unitarity. The L+P approach 
separates pole and regular parts in the form of a Laurent expansion, and 
instead of modeling the regular part using some physical model it uses the 
conformal-mapping-generated, rapidly converging power series with 
well-defined analytic properties called a Pietarinen expansion to represent 
it effectively. In other words, the method replaces the regular part, 
calculated in a model with the simplest analytic function that has the 
correct analytic properties of the analyzed partial wave (multipole),
and which fits the given input.  In such an approach, the model dependence 
is minimized, and is reduced to the choice of the number and location of 
L+P branch-points used in the model.

So, we expand the multipoles, $M(W)$, in terms of a sum over all poles and
with a Pietarinen series representing the  energy-dependent regular (non-pole) 
part as:
\begin{eqnarray}
        M(W) & = & \sum_{i=1}^{k}\frac{\alpha_{-1}^{(i)}}{W - W_i} + B^L(W) .
        \label{eq:LP}
\end{eqnarray}\noindent
Here $W$, $\alpha_{-1}^{(i)}$, and $W_i$ are complex numbers representing 
the c.m.\ energy, residues, and pole positions for the \textit{i}th pole, 
respectively, and $B^L(W)$ is a regular function in the whole complex plane. 
As shown in Ref.~\cite{Svarc:2013laa}, a general unknown analytic 
function $B(W)$ with branch-points in $x_P$, $x_Q$, and $x_R$ can be expanded 
into a power series of Pietarinen functions as
\begin{eqnarray} \label{L+P}
        B^L(W)& = & \sum_{n=0}^{M}c_nX(W)^n +  \sum_{n=0}^{N}d_nY(W)^n
        + \sum_{n=0}^{N}e_nZ(W)^n +..., \nonumber \\
        &  &  X(W) =  \frac{\alpha^2 - \sqrt{x_P - W}}{\alpha^2 + \sqrt{x_P - W}} , \nonumber \\
        &  &  Y(W) =  \frac{\beta^2  - \sqrt{x_Q - W}}{\beta^2  + \sqrt{x_Q - W}} , \nonumber \\
        &  &  Z(W) =  \frac{\gamma^2 - \sqrt{x_R - W}}{\gamma^2 + \sqrt{x_R - W}}
\end{eqnarray}\noindent
where $c_n$, $d_n$, $e_n$ and $\alpha$, $\beta$, $\gamma$ are real numbers that 
represent tuning parameters and coefficients of the Pietarinen functions $X(W)$, 
$Y(W)$, and $Z(W)$, respectively. A variable number of coefficients in three series 
of Eq.~(\ref{L+P}) were used, depending on the structure of the non-pole part of 
each amplitude.

When the input data statistically satisfies the normal distribution (meaning that 
they are acquired in a non-correlated procedure), the estimation of the errors 
of all resulting pole parameters can be obtained directly from any standard 
minimization routine.  However, in our case, the nearby energy points of the 
input multipoles are correlated through analyticity of the energy-dependent 
partial wave of MA19 solution, so the standard error analysis cannot be used as 
the standardly defined $\chi^2$ becomes extremely small ($\chi^2 << 1$)
regardless of which error is attributed to the input. So, the error analysis of 
resulting parameters cannot be reliably performed.

In this paper, we improve this aspect of our model, and introduce a procedure 
that creates completely realistic errors for pole parameters extracted from
ED MA19 partial waves. First, we have to attribute realistic errors to MA19 ED 
solutions. We do it by using MA19 single-energy (SE) analysis as the measure of 
how good an ED analysis actually is. We perform SE PWA at energies where we have 
an abundance of experimental data, and constrain it to an ED MA19 partial wave 
strong enough to achieve the continuity in energy, and weak enough to give 
enough freedom for the fit to move away as much as possible from the ED MA19 
solution in coming maximally close to the experiment.  In that way at each 
energy the SE solution maximally reproduces the available experimental data base, 
so we are definitely closer to experiment than the ED MA19 solution is.  The 
probability that the true value lies inside the interval which is defined by 
the difference between partial wave values in the ED and SE points \mbox{dif  
=  $PW_{ED}(W) -PW_{SE}(W)$} is, therefore, close to  100 \%, so we define the 
standard deviation of the partial wave ED value as $\sigma_{PW} = {\rm dif}/3$.

The next step is eliminating the correlations between neighboring energy points 
in the ED PW, which is introduced by the analyticity of the fitting function. 
This is done by randomizing central values of the ED solution with PW standard
deviation $\sigma_{PW}$, and assigning the error of the source ED 
error of the randomized point.  In this way, we generate 1000 ensembles of 
randomized ED, which then independently fitted, and an ensemble of 
1000 pole parameters for the investigated partial waves was obtained. We 
confirm that  the obtained ensemble corresponds to the normal distribution by 
generating the probability density function of the ensemble, and verifying that 
the shape of the obtained histogram is well reproduced with this, properly 
normalized, function. If this is the case, we then make a standard normal 
distribution error analysis of the generated ensemble, and pole parameters with 
realistic errors are obtained. In cases where this criterion is not matched, we 
have to modify the obtained ensemble by cutting out the points which erroneously 
enter this ensemble, and which belong to the nearby local minimum of the L+P 
minimization procedure. When the new ensemble matches the criterion, we are 
free to make the desired error analysis.  Once the pole position and residue 
were determined, the photon decay amplitude at the pole could be constructed, 
as described in Ref.~\cite{Workman:2013rca}. The residue of the corresponding 
$\pi N$ elastic scattering amplitude, required in this construction, was taken 
from the SAID analysis of elastic scattering data~\cite{Arndt:2006bf}. The 
spread of determinations found in Ref.~\cite{Tanabashi:2018oca} was used to 
estimate its uncertainty.

Final results for the photon-decay amplitudes are listed in Table~\ref{tab:tbl3}. 
Here comparisons are made with the Bonn-Gatchina (BnGa) and Kent State University 
(KSU) values and with an 
earlier SAID determination. For the $N(1520)$, the PDG2018 lists only BW values. 
This being the first determination of pole values, we compared at the level of 
moduli, finding good agreement. The agreement between BW and pole values is not 
as good for the Roper resonance, where the complicated pole-cut structure may 
invalidate this simple comparison of pole and BW quantities.
\begin{table}[htb!]

\centering \protect\caption{Moduli [in (GeV)$^{-1/2}\times 10^{-3}$] and 
	phases (in degrees) of the photon decay amplitudes at the pole 
	for neutron A$_{1/2}$(n) and A$_{3/2}$(n) from the SAID 
	MA27~\protect\cite{Mattione:2017fxc} and MA19 solutions. Pole
        results from the Bonn-Gatchina (BnGa) analysis are included for
        comparison~\protect\cite{Anisovich:2017xqg} [BW values are
        from Ref.~\protect\cite{Anisovich:2013jya}].  Kent State Univ. (KSU) 
	results are from Ref.~\protect\cite{Hunt:2018wqz}. BW values labeled 
	with a $\dagger$. }
\vspace{2mm}
{\begin{tabular}{|c|c|c|c|c|c|} 
\hline
Resonance      & Coupling     &           SAID MA19           &         SAID MA27            &        BnGa                  & KSU \tabularnewline
\hline
N(1440)1/2$^+$ & A$_{1/2}$(n) &  80$\pm$10,  9$\pm$2$^\circ$  & 65$\pm$5,   5$\pm$3$^\circ$  &  43$\pm$12$^\dagger$         & 13$\pm$12$^\dagger$ \tabularnewline
N(1520)3/2$^-$ & A$_{3/2}$(n) &-130$\pm$8,  20$\pm$6$^\circ$  &                              &-113$\pm$12$^\dagger$         &-123$\pm$6$^\dagger$ \tabularnewline
N(1520)1/2$^-$ & A$_{1/2}$(n) & -47$\pm$4,   1$\pm$2$^\circ$  &                              & -49$\pm$8$^\dagger$          & -72$\pm$3$^\dagger$ \tabularnewline
N(1535)1/2$^-$ & A$_{1/2}$(n) & -70$\pm$10,  2$\pm$5$^\circ$  &-55$\pm$5,   5$\pm$2$^\circ$  & -88$\pm$4, 5$\pm$4$^\circ$   & -55$\pm$6$^\dagger$ \tabularnewline
N(1650)1/2$^-$ & A$_{1/2}$(n) &  13$\pm$4, -50$\pm$15$^\circ$ & 14$\pm$2, -30$\pm$10$^\circ$ &  16$\pm$4, -28$\pm$10$^\circ$&   1$\pm$6$^\dagger$  \tabularnewline
\hline
\end{tabular}} \label{tab:tbl3}
\end{table}

\section{Summary and Conclusion}
\label{sec:sum}

The $\gamma n\rightarrow\pi^0n$ differential cross sections 
have been measured at the tagged photon facility of the Mainz 
Microtron MAMI using the Crystal Ball spectrometer. The data 
span the photon-energy range 290--813~MeV ($W$ = 1.195 -- 
1.553~GeV) and from 18$^\circ$ to 162$^\circ$ c.m.\ angular 
range. The accumulation of $3.6\times 10^6$ $\gamma 
n\rightarrow\pi^0n$ events allows fine binning of the data in 
energy and angle, which will enable the reaction dynamics to 
be studied in greater detail.  The present differential cross 
section data are in reasonable agreement with the previous 
measurements, but the energy binning much finer, we covered 
much low energies and something for the very low energy
range in the maximum of the $\rm\Delta$-isobar, and we used 
FSI corrections.  Additionally, the total photoabsorption 
cross section was measured. 

Differential cross sections have visible discrepancies when
compared to the predictions from SAID, MAID, and BnGa at 
energies below 300~MeV and are in satisfactory agreement with 
PWA results at higher-energies. 

A comparison of determinations for photon-decay amplitudes at 
the pole shows reasonably good agreement. The only noticeable 
exception is seen in the Roper photon-decay amplitude.  Here,
the difference may be due to the comparison of BW
and pole-valued quantities. It will be interesting to see
updated BnGa results at the pole once these data have been
incorporated into new fits.


\section{Acknowledgements}

This work was supported in part by the U.S. Department of Energy, 
Office of Science, Office of Nuclear Physics, under Awards No. 
DE--FG02--01ER41194, DE--SC0016583, DE--SC0016582, and 
DE--SC0014323. We would like to thank all the technical and 
non-technical staff of MAMI for their support. This work was 
supported by Schweizerischer Nationalfonds (Grant No. 200020- 
132799, 121781, 117601, 113511), Deutsche Forschungsgemeinschaft 
(SFB Grant No. 443, SFB/TR 16, SFB 1044), DFG-RFBR (Grant No. 
05--02--04014), UK Science and Techology Facilities Council 
(STFC Grant No. 57071/1, 50727/1), Russian Foundation for Basic 
Research (RFBR Grant No. 16--02--00767--a), European Community 
Research Infrastructure Activity (FP6), the U.~S. DOE, U.~S. 
NSF, and NSERC (Grant No. SAPPJ--2018--00020) Canada.


\end{document}